\documentclass[11pt]{article}
\usepackage{mint,psfig}

\title{
\vspace{-1cm}
\begin{flushright}
\normalsize     YITP-01-53 \\
               UTCCP-P-108 \\
              RCNP-Th01017
\end{flushright}
\vspace{1cm}
Numerical study of $O(a)$ improved Wilson quark action \\
on anisotropic lattice}

\author{
Hideo~Matsufuru$^1$, Tetsuya~Onogi$^1$ and Takashi~Umeda$^2$
   \bigskip \\
 {\it $^1$Yukawa Institute for Theoretical Physics, Kyoto University,}\\
            {\it Kyoto 606-8502, Japan}
   \smallskip \\
 {\it $^2$Center for Computational Physics, University of Tsukuba,}\\
            {\it Tsukuba 305-8577, Japan}
}

\date{30 June 2001}

\begin{document}

\maketitle

\begin{abstract}
The $O(a)$ improved Wilson quark action on the anisotropic lattice
is investigated.
We carry out numerical simulations in the quenched approximation
at three values of lattice spacing ($a_{\sigma}^{-1}=1$--$2$ GeV)
with the anisotropy $\xi=a_{\sigma}/a_{\tau}=4$,
where $a_{\sigma}$ and $a_{\tau}$ are the spatial and the temporal
lattice spacings, respectively.
The bare anisotropy $\gamma_F$ in the quark field action is
numerically tuned by the dispersion relation of mesons so that
the renormalized fermionic
anisotropy coincides with that of gauge field.
This calibration of bare anisotropy is performed to
the level of 1 \% statistical accuracy in the quark mass region
below the charm quark mass.
The systematic uncertainty in the calibration is estimated
by comparing the results from different types of dispersion relations,
which results in 3 \% on our coarsest lattice and tends to
vanish in the continuum limit. In the chiral limit, there is an 
additional systematic uncertainty of 1 \%
from the chiral extrapolation.

Taking the central value $\gamma_F=\gamma_F^*$ from the result 
of the calibration, we compute the light hadron spectrum.
Our hadron spectrum is consistent with the result by UKQCD
Collaboration on the 
isotropic lattice. We also study the response of the hadron 
spectrum to the change of anisotropic parameter,
$\gamma_F \rightarrow \gamma_F^* + \delta\gamma_F$.
We find that the change of $\gamma_F$ by 2 \% induces a change of 1 \%
in the spectrum for physical quark masses.
Thus the systematic uncertainty on the anisotropic lattice,
as well as the statistical one, is under control.
\end{abstract}

\section{Introduction}
  \label{sec:introduction}

The anisotropic lattice is drawing more attention as a 
useful technique of the lattice QCD simulation in various 
physics such as the spectroscopy of exotic states,  
finite temperature QCD and the heavy quark physics.
However, the advantage of having a fine lattice spacing 
in the temporal direction is obtained at the sacrifice of
manifest temporal-spatial axis-interchange symmetry.
Therefore improper use of the anisotropic lattice 
could lead to an unphysical result due to the lack of Lorentz symmetry 
in the continuum limit. This can be a serious problem for those
physics in which the precision of the results is crucial. 

One way to avoid this problem is to tune the anisotropy 
parameters of the action by imposing the conditions
with which the Lorentz invariance is satisfied for some
physical observables.
In the Wilson plaquette gauge action there is only one anisotropic
parameter in the action \cite{Kar82}.
Wilson loops are used to obtain the
relation between the anisotropic parameter and the physical ratio of 
the lattice spacings in temporal and spatial directions,
$\xi=a_{\sigma}/a_{\tau}$, where $a_{\sigma}$ and $a_{\tau}$
are the spatial and the temporal lattice spacings
\cite{Bur88,TARO97,KES98,Kla98}.
On the other hand, not much is known about the quark action.
This is because the previous works with the quark action on the
anisotropic lattice have been devoted to the charmonium systems
\cite{Ume01,Kla99,Chen00,CPPACS01}.
In order to apply the anisotropic lattice to the systems containing
the light quarks, one has to study how one can tune the parameter 
of light quark action in practical simulations.

In this paper we study the $O(a)$ improved Wilson action on 
the anisotropic lattice
using the quenched lattices with three lattice spacings, at fixed
renormalized anisotropy, $\xi=4$. These scales covers the range of 
the spatial lattice cutoff $a_{\sigma}^{-1}=$1--2 GeV.
We first tune the bare anisotropy in the quark action numerically
so that the renormalized fermionic anisotropy is equal to that
of the gauge field by imposing the relativistic dispersion relation
of mesons.
This calibration is performed to
the level of 1 \% statistical accuracy in whole quark mass region
below the charm quark mass.
The extrapolation of tuned bare anisotropy to the chiral limit
is performed by fitting to presumable forms, and causes additional
systematic uncertainty of 1 \% at the chiral limit.
The systematic uncertainty in the calibration from the lattice
artifact is estimated
by comparing the results from different types of dispersion relations.
It results in 3 \% on our coarsest lattice, and tends to
vanish in the continuum limit.
Using the result of the calibration we compute the light hadron
spectrum on anisotropic lattices, and examine how
the uncertainty in the calibration affect the spectrum at the
parameters of physical interest.
It is found that the uncertainty of 2 \% in calibration induces
systematic error of 1 \% in the spectrum.
Thus the systematic uncertainty on the anisotropic lattice
is under control, as well as the statistical one.
We also show that the anisotropic lattices produce consistent results 
with those on the isotropic lattices, and that the Lorentz 
invariance of the simple matrix element is satisfied within errors. 

This paper is organized as follows.
The next section describes the $O(a)$ improved Wilson quark action,
which has been discussed in \cite{Aniso01a,Ume01}.
The calibration procedures are discussed in 
Section~\ref{sec:calibration}.
In Section~\ref{sec:numerical}, we perform the calibration of
the bare anisotropy in the quark action.
The systematic uncertainty induced by this tuning is fully
examined.
In Section \ref{sec:spectroscopy} we apply the anisotropic lattice
to the light hadron spectroscopy.
In the last part of this section, the systematic
uncertainty due to the anisotropy is again investigated in terms
of the effect on hadron spectrum.
The last section is devoted to our conclusions.

\section{Quark action on the anisotropic lattice}
\label{sec:formulation}

\subsection{Quark field action}

We employ the $O(a)$ improved quark action on the anisotropic lattice.
The form of action has been discussed in Ref.~\cite{Aniso01a},
which is the same as the Fermilab action \cite{EKM97}
but defined on the anisotropic lattice.
In this section we summarize the result which will be necessary 
in the following calculations. 

The quark action is represented in the hopping parameter form as
\begin{equation}
 S_F = \sum_{x,y} \bar{\psi}(x) K(x,y) \psi(y),
\end{equation}
\begin{eqnarray}
 K(x,y) &=& \delta_{x,y}
     - \kappa_{\tau} \left[ (1-\gamma_4)U_4(x)\delta_{x+\hat{4},y}
        + (1+\gamma_4)U_4^{\dag}(x-\hat{4})\delta_{x-\hat{4},y} \right]
 \nonumber \\
 & & \hspace{0.7cm}
    - \kappa_{\sigma} \sum_{i}
      \left[ (r-\gamma_i)U_i(x)\delta_{x+\hat{i},y}
      + (r+\gamma_i)U_i^{\dag}(x-\hat{i})\delta_{x-\hat{i},y} \right]
 \nonumber \\
 & & \hspace{0.7cm}
    -  \kappa_{\sigma} c_E
              \sum_{i} \sigma_{4i}F_{4i}(x)\delta_{x,y}
    - r \kappa_{\sigma} c_B
              \sum_{i>j} \sigma_{ij}F_{ij}(x)\delta_{x,y},
 \label{eq:action}
\end{eqnarray}
where $\kappa_{\sigma}$ and  $\kappa_{\tau}$ 
are the spatial and the temporal hopping parameters,
$r$ is the Wilson parameter
and  $c_E$ and $c_B$ are  the clover coefficients.
In principle for a given $\kappa_{\sigma}$, the 
four parameters $\kappa_{\sigma}/\kappa_{\tau}$, $r$, $c_E$ and $c_B$
should be tuned so that Lorentz symmetry holds up to 
discretization errors of $O(a^2)$.

On the anisotropic lattice, the mean-filed values of the spatial 
link variable $u_{\sigma}$ and the temporal one $u_{\tau}$ are
different from each other.
The tadpole-improvement \cite{LM93}
is achieved by rescaling the link variable
as $U_i(x) \rightarrow U_i(x)/u_{\sigma}$ and  $U_4(x) \rightarrow
U_4(x)/u_{\tau}$. This is equivalent to redefine the
hopping parameters with the tadpole-improved ones (with tilde)
through $\kappa_{\sigma} = \tilde{\kappa}_{\sigma}/u_{\sigma}$
and $\kappa_{\tau} = \tilde{\kappa}_{\tau}/u_{\tau}$.
We define the anisotropy parameter $\gamma_F$ as
\begin{equation}
\gamma_F
  \equiv \frac{\tilde{\kappa}_{\tau}}{\tilde{\kappa}_{\sigma}}.
\end{equation}
At the tadpole-improved tree-level, and for sufficiently small quark
mass, the anisotropy $\gamma_F$
coincides with the cutoff anisotropy $\xi=a_{\sigma}/a_{\tau}$.

In this work, we set the coefficients of the spatial part of the Wilson
term and the clover coefficients as the tadpole-improved tree-level
values, namely,
\begin{equation}
 r = \frac{1}{\xi}, \hspace{1cm}
 c_E=\frac{1}{u_{\sigma} u_{\tau}^2}, \hspace{1cm}
 c_B=\frac{1}{u_{\sigma}^3}
\label{eq:cecb}
\end{equation}
and perform a nonperturbative calibration only for $\gamma_F$ 
with the meson dispersion relation.

It is useful to define $\kappa$ 
\begin{eqnarray}
\frac{1}{\kappa} \equiv \frac{1}{\tilde{\kappa}_{\sigma}}
     - 2(\gamma_F+3r-4)
 \label{eq:kappa}
\end{eqnarray}
so that the bare quark mass in the spatial lattice unit, $m_{0\sigma}$, is
expressed as
\begin{eqnarray}
   m_{0\sigma} = \frac{1}{2} \left( \frac{1}{\kappa} - 8 \right),
\end{eqnarray}
which is analogous to the case of the isotropic lattice.

\subsection{Dispersion relation of free quark}
\label{subsec:free_quark}

In this subsection, we examine the dispersion relation of the free
quark on the anisotropic lattice.
First the tree-level relation of bare anisotropy with $\xi$
is derived from the condition that the rest mass and the kinetic
mass coincide.
Then, we discuss how the dispersion relation is distorted at
the edge of Brillouin zone due to our choice, $r=1/\xi$
\cite{Ume01}.

From the action (\ref{eq:action}), the free quark propagator satisfies
the dispersion relation
\begin{equation}
\cosh E(\vec{p}) = 1 + \frac{\vec{\bar{p}}^2
            + (m_0 + \frac{1}{2} \frac{r}{\gamma_F} \vec{\hat{p}}^2 )^2}
            { 2 (1+m_0 + \frac{1}{2} \frac{r}{\gamma_F} \vec{\hat{p}}^2 ) },
\label{eq:dispersion}
\end{equation}
where $\bar{p}_i = \frac{1}{\gamma_F}\sin p_i$,
$\hat{p}_i=2\sin(p_i/2)$ and $m_0=m_{0\sigma}/\gamma_F$ is the
bare quark mass in the temporal lattice unit.
Setting $\vec{p}=0$, Eq. (\ref{eq:dispersion}) gives the rest mass
\begin{equation}
 M_1 \equiv E(\vec{0}) = \ln (1+m_0).
\label{eq:rest_mass}
\end{equation}
On the other hand, the kinetic mass is defined and obtained as
\begin{equation}
 \frac{1}{M_2}
   \equiv \left. \xi^2 \frac{d^2E}{dp_i^{\,2}} \right|_{\vec{p}=0}
  = \xi^2 \left(  \frac{\frac{r}{\gamma_F}}{m_0+1} 
                + \frac{\frac{2}{\gamma_F^2}}{m_0(m_0+2)} \right).
\label{eq:kin_mass}
\end{equation}
Generally the rest mass (\ref{eq:rest_mass}) and the
kinetic mass (\ref{eq:kin_mass}) are different.
One can tune the bare anisotropy parameter $\gamma_F$
so that it gives the same values for the rest and the
kinetic masses \cite{EKM97}.
Putting the rest and the kinetic masses equal, the anisotropy
parameter $\gamma_F$ is represented with $m_0$ and $r$ as
\begin{equation}
\frac{1}{\gamma_F}
 = \sqrt{  \left( \frac{rm_0(m_0+2)}{4(m_0+1)} \right)^2
                 + \frac{m_0(m_0+2)}{2\xi^2\ln(1+m_0)} } 
 - \frac{r m_0(m_0+2)}{4(m_0+1)}.
\end{equation}
For small $m_0$, $\gamma_F$ is expanded in $m_0$ as
\begin{eqnarray}
\frac{1}{\gamma_F}
 &=& \frac{1}{\xi}\left[ 1 + \frac{1}{2}(1-r\xi)m_0
     + \frac{1}{24} (-1 + 6 r \xi + 3 r^2 \xi^2 ) m_0^2 \right] 
 \nonumber \\
 &=& \frac{1}{\xi}\left[ 1 + \frac{1}{3} m_0^2 \right] 
    \hspace{1cm} (r=1/\xi).
\label{eq:gamma_F_in_m_0}
\end{eqnarray}
The $m_0$ dependence starts with the quadratic term for $r=1/\xi$,
therefore the dependence on the quark mass is small for sufficiently
small $m_0$.
For example, let us consider the case of $a_{\tau}=4$ GeV,
which corresponds to our coarsest lattice in the simulation.
The charm quark mass corresponds to $m_0\simeq 0.3$ and at this
value $\gamma_F$ is different from $\xi$ only 3 \%.
Up to this quark mass region, one can expect that the difference of
$\gamma_F$ from $\xi$ is also small in the numerical simulation.
This is examined in Section~\ref{sec:numerical}.

\begin{figure}[tb]
\centerline{\psfig{file=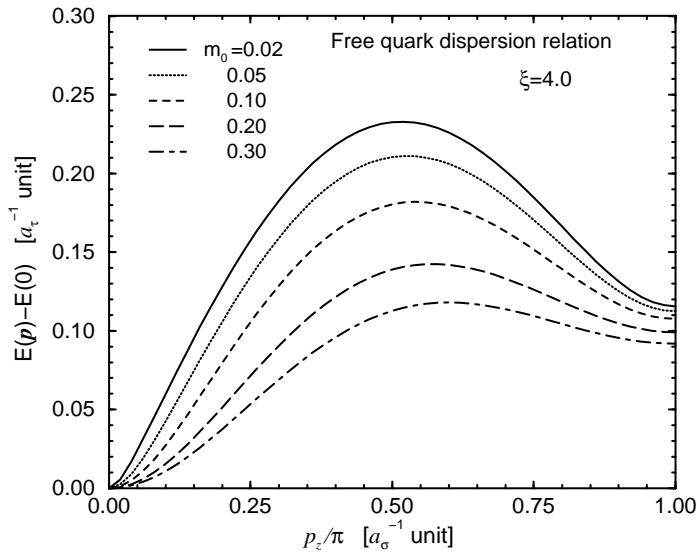,width=0.8\figwidth}}
\caption{Dispersion relation of the free quark for $\xi=4$.
}
\label{fig:dispersion}
\end{figure}

With our choice of Wilson parameter, $r=1/\xi$,
the action (\ref{eq:action}) leads the smaller spatial Wilson term
for the larger cutoff anisotropy $\xi$.
The question is how the contribution of the doubler eliminated
by the Wilson term becomes significant for practical value of $\xi$.
In the following argument on this subject,
we only treat the case $\gamma_F$ almost equal
to $\xi$, i.e. the region of $m_0$ sufficiently
smaller than unity.
Figure~\ref{fig:dispersion} shows the dispersion relation
(\ref{eq:dispersion}) for several values of $m_0$
in the case of $\xi=4$, which we use in the numerical simulation.

Let us examine the practical case, $a_{\sigma}^{-1}\simeq 1.0$ GeV,
which corresponds to the lowest spatial cutoff of our three lattices.
For the light quark mass region,
$m_0=0.02$--$0.05$ corresponds to 80--200 MeV, and roughly covers
the mass region which we use in the hadron spectroscopy
in Section~\ref{sec:spectroscopy}.
$E(\vec{p})-E(0)$ rapidly decrease at the edge of the Brillouin zone,
and the height at $z=a/\pi$ is around 400 MeV.
For two quarks with momenta $p=\pm a/\pi$, additional energy of
doublers is $\sim 800$ MeV,
and is expected to affect not severely on the spectrum and other
observables.
For higher lattice cutoff, the situation becomes better.
Then we regard the doubler contribution is sufficiently small on
the lattice we use in the simulation.
$m_0=0.3$ roughly corresponds to the charm quark mass with 
$a_{\sigma}^{-1}\simeq 1$ GeV.
In the case of heavy-light hadrons, such as $D$ mesons and
$\Lambda_c$ baryon, the scale of momentum transfered inside hadrons
is of the order of $\Lambda_{QCD}$, and the same argument for
light quark case holds.
On the other hand, for the heavy quarkonium, 
the typical energy and momentum exchanged inside the meson are
in the order of $mv^2$ and $mv$ respectively \cite{TL91}.
For the charmonium, $v^2 \sim 0.3$, then typical scale
of the kinetic energy is around $500$ MeV.
It seems not sufficiently smaller than the two doublers' contribution,
and hence one need to choose larger lattice cutoff
in the calculation of heavy quarkonium.

\section{Calibration procedures}
\label{sec:calibration}

On anisotropic lattice, one must tune the parameters so that the
anisotropy of quark field, $\xi_F$, equals to that of the gauge
field, $\xi_G$:
\begin{equation}
 \xi_F(\beta,\gamma_G;\kappa,\gamma_F)
 = \xi_G(\beta,\gamma_G;\kappa,\gamma_F)
= \xi
\label{eq:xi}
\end{equation} 
Since $\xi_G$ and $\xi_F$ are in general functions of both of
gauge parameters ($\beta$, $\gamma_G$) and quark parameters,
($\kappa$, $\gamma_F$),
the simulation with dynamical quarks requires to tune
these bare parameters simultaneously.
In the quenched case, however, this tuning is rather easy to
be performed, since $\xi_G$ can be determined independently of
$\gamma_F$.
After the determination of $\xi$, one can tune
$\gamma_F$ so that a certain observable satisfies the condition
(\ref{eq:xi}).
In this work, we use the relativistic dispersion relation of
meson,
\begin{equation}
E^2 (\vec{p}) = m^2 + \frac{\vec{p}^2}{\xi_F^2} + O(\vec{p}^4),
\label{eq:DR1}
\end{equation}
as our main calibration procedure.
The energy and the mass of meson, $E$ and $m$, are in the temporal
lattice unit while the momentum is in the spatial lattice unit.
The $\xi_F$ appears to convert the momentum in the spatial lattice
unit into the temporal lattice unit, and considered as the fermionic
anisotropy defined through this relation.
With the condition $\xi_F=\xi$, this condition satisfies that
the rest mass and the kinetic mass equal to each other.
For finite lattice spacings, above dispersion relation only holds
up to the $O((\vec{p}^2)^2)$ correction term.
In the continuum limit, this higher order term in $a$ would
vanish and the relativistic dispersion relation would be restored.

In the numerical simulation, we fit $E^2$ to the form Eq.~(\ref{eq:DR1})
and obtain the value of $\xi_F$ for each input value of bare anisotropy
$\gamma_F$.
Then we linearly interpolate $\xi_F$ in terms of $\gamma_F$
and find out $\gamma_F^*$, the value of $\gamma_F$ with which 
$\xi_F=\xi$ holds.

In order to estimate the systematic errors we also 
use the dispersion relation which corresponds to 
the lattice Klein-Gordon action \cite{Ume01},
\begin{equation}
\cosh E(\vec{p}) - \cosh E(\vec{p}=0)
   = 1 + \frac{1}{2\xi_{KG}^2} \vec{\hat{p}}^2.
\label{eq:DR2}
\end{equation}
Thus the comparison of these two calibration conditions typically
shows the size of the lattice discretization errors.

Expanding this expression in $a$, 
$\xi_{KG}$ is related to $\xi_F$ as
\begin{equation}
  \xi_{KG} = \xi_F \left( 1 - \frac{m^2}{12} +O(a^4) \right).
\label{eq:DR3}
\end{equation}
The same input $\gamma_F$ gives smaller value for
$\xi_{KG}$ than $\xi_F$, and therefore the tuned bare anisotropy
$\gamma_F^{*}$ results in larger value in the former case.

\section{Numerical Results of Calibration}
\label{sec:numerical}

The goal of this section is to determine the tuned bare
anisotropy of quark field, $\gamma_F^*$, at each fixed quark mass
in the region from strange to charm quark masses.
The reason for this choice of the quark mass range is that 
the simulation is easier which saves the amount of work in 
the exhaustive study for calibrations. 
Fitting the result as a function of the quark mass, we obtain 
$\gamma_F^*$ to the statistical accuracy of 1 \% level for the whole
quark mass region below the charm quark mass including the chiral limit.

Then we estimate the systematic uncertainties of  $\gamma_F^*$
which are mainly due to $O(\alpha a)$ and  $O(a^2)$ lattice artifacts.
We also investigate how these systematic errors as well as the
statistical error affect the meson masses, in the region 
$m_s < m_q < m_c$.
The response of hadron masses with respect to $\gamma_F$ in
the light quark mass region, $ m_q < m_s$, needs additional care,
and is the subject of next section.
At the end of this section, we summarize the result of calibration.

\subsection{Simulation Parameters for the Calibration}

In this work, we use three lattices with $\beta=5.75$, $5.95$ and
$6.10$ and the renormalized anisotropy $\xi=4$.
The value of $\gamma_G$ corresponding to the desired value of $\xi$
has been studied in detail by Klassen \cite{Kla98},
and we can use his relation of $\gamma_G$ and $\xi$
which is obtained in one percent accuracy.
The statistical uncertainties are, otherwise noted, estimated by
the single elimination Jackknife method with appropriate binning.
The configurations are separated by 2000 (1000) pseudo-heat-bath sweeps,
after 20000 (10000) thermalization sweeps at $\beta$=5.95 and 6.10
(5.75).
The configurations are fixed to the Coulomb gauge,
which is particularly useful for the smearing of hadron operators.

The lattice cutoffs and the mean-field values of link variables
are determined on the smaller lattices with half size in temporal
extent for $\beta=5.75$, $5.95$, and otherwise with the same
parameters, while at $\beta=6.10$ the lattice size is $16^3\times 64$.
To obtain the lattice cutoffs, the static quark potential is
measured with standard procedure.
We adopt the hadronic radius $r_0$ proposed by Sommer \cite{Som94}
to set the scale.
Following the method in Ref.~\cite{Som94}, 
we determine the force between static
quark and antiquark, as a function of $r_I$, the interquark distance
improved with the lattice one-gluon exchange potential form.
Then we fit the values of force, containing the off-axis data, 
to the form $\sigma + A/r_I^2$ in the fitting region roughly
$0.5 r_0 < r_I < 2 r_0$.
The parameters $\sigma$ and $A$ would be identified as the string
tension and the Coulomb coefficient.
The systematic uncertainty due to the choice of fit range is
small, and at most the same size as the statistical error.
Table~\ref{tab:parameters} shows the value of $r_0$ and
the $a_{\sigma}^{-1}$ determined by setting the physical value of
$r_0$ as $r_0^{-1}=395$ MeV ($r_0\simeq 0.5$ fm).
The quoted error represent only the statistical uncertainty.

The mean-field values, $u_{\sigma}$ and $u_{\tau}$, are obtained
as the average of the link variables in the Landau gauge,
where the mean-field values are used
self-consistently in the fixing condition \cite{Ume01}.
The results are also listed in Table~\ref{tab:parameters}.
The mean-field value of the temporal gauge field has small error,
and close to unity.
$\eta_{MF}=u_{\tau}/u_{\sigma}$, the mean-field estimate of
$\eta=\xi/\gamma_G$, is close to the
value of $\eta$ determined nonperturbatively by Klassen.
This suggest that the tadpole-improvement works well also on
the anisotropic lattice.

\begin{table}[tb]
\caption{
Lattice parameters. The scale $a_{\sigma}^{-1}$ is determined from
the hadronic radius $r_0$.
The mean-field values are in the Landau gauge.
The statistical uncertainty of  $u_{\tau}$ is less than the last digit.}
\begin{center}
\begin{tabular}{cccccccc}
\hline\hline
$\beta$ & $\gamma_G$ & size & $r_0$ &  $a_{\sigma}^{-1}$ [GeV] &
 $u_{\sigma}$ & $u_{\tau}$ & $\eta_{MF}$ \\
\hline
5.75 & 3.072~ & $12^3\times ~96$ & 2.786(15)& 1.100( 6)&
      0.7620(2) & 0.9871 & 1.2953(4) \\
5.95 & 3.1586& $16^3\times 128$ & 4.110(23)& 1.623( 9)&
      0.7917(1) & 0.9891 & 1.2494(2) \\
6.10 & 3.2108& $20^3\times 160$ & 5.140(32)& 2.030(13)&
      0.8059(1) & 0.9901 & 1.2285(2) \\
\hline\hline
\end{tabular}
\end{center}
\label{tab:parameters}
\end{table}

\subsection{Quark field calibration}

As described in Section~\ref{sec:calibration}, we use the
relativistic dispersion relation in the calibration of
parameters in the quark action.
Since the gauge field calibration is in the accuracy of one percent,
we aim to tune the quark parameters to a similar level.

For convenience, we choose $\kappa$ and $\gamma_F$
as the input parameters and determine $\kappa_{\sigma}$
and $\kappa_{\tau}$ along eq.~(\ref{eq:kappa}).
Fixing $\kappa$ corresponds to fixing
the bare quark mass in the spatial lattice unit.
For each values of $(\kappa, \gamma_F)$, the pseudoscalar
and vector meson correlators are obtained with zero and finite 
momenta.
The fermionic anisotropy $\xi_F$ is defined through the
relativistic dispersion relation, eq.~(\ref{eq:DR1}).
We assume the linear dependence of $\xi_F$ on $\gamma_F$ in the
vicinity of $\xi_F \simeq \xi$. 
We use linear interpolation to 
obtain $\gamma_F^*$, the value of $\gamma_F$ with which
the relation $\xi_F= \xi$ holds.

\paragraph{Result of the dispersion relation.}

The parameters $(\kappa,\gamma_F)$ used in the calibration are
listed in Table~\ref{tab:qprm_E}, \ref{tab:qprm_F} and
\ref{tab:qprm_G} for $\beta=5.75$, $5.95$ and $6.10$, respectively.
As the meson operators at the source, we adopt the smeared
operators with appropriate smearing functions.
For the light quark region, the Gaussian function is used as the
smearing function, with the width of $0.2 \sim 0.4$ fm.
In the charm quark mass region, we also use the measured wave function 
for the smearing function.
We measure the two point functions for momenta 
$\vec{p}=\vec{n}\cdot (2\pi/L)$, where
$L$ is spatial lattice extent and $\vec{n}=(0,0,0)$, $(1,0,0)$, $(1,1,0)$,
$(1,1,1)$ and $(2,0,0)$.
All rotationally equivalent $\vec{n}$'s are averaged.
The standard procedure is used in extracting the energy at each
momentum.

The energies are then fitted to the linear or
quadratic forms in $\vec{p}^2$ to extract the fermionic anisotropy
$\xi_F$ in each channel.
In the case of linear fit, we use only three lowest momentum states,
$\vec{n}=(0,0,0)$, $(1,0,0)$ and $(1,1,0)$.
We assume the linear dependence of $\xi_F(\gamma_F)$
in $\gamma_F$, and this is indeed verified in several
examples.

\begin{figure}[tb]
\centerline{\psfig{file=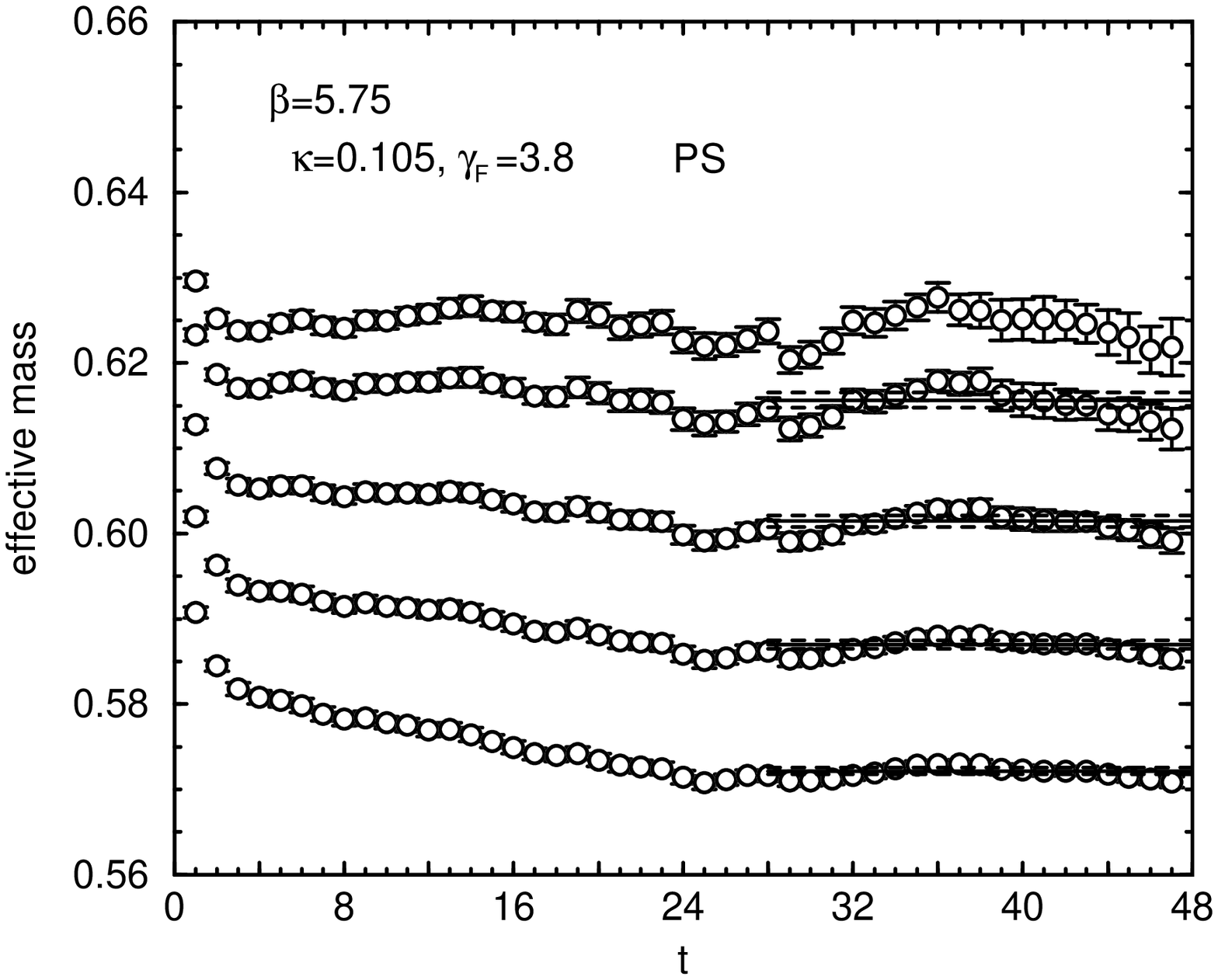,width=0.69\figwidth}
\hspace{-1.2cm}
\psfig{file=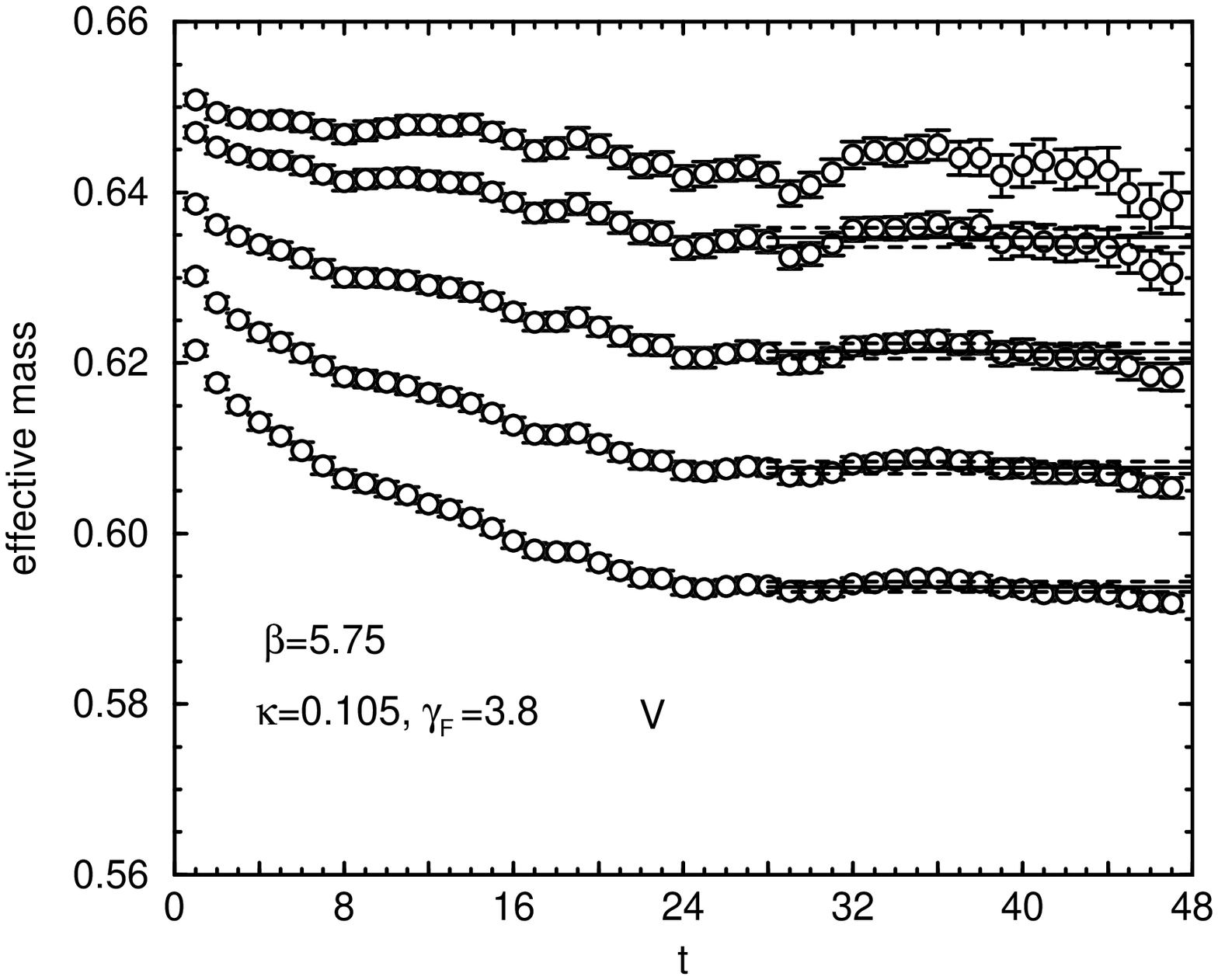,width=0.69\figwidth}}
\vspace{-0.6cm}
\caption{
Effective mass plots of PS and V mesons at $\kappa=0.105$,
$\gamma_F=3.8$ on $\beta=5.75$ lattice.
From bottom to top, the states with integer momentum vectors $\vec{n}=(0,0,0)$,
$(1,0,0)$, $(1,1,0)$, $(1,1,1)$ and $(2,0,0)$.
Horizontal solid lines represent the result of fit of correlators
and the fitting range. 
The statistical errors are represented by the dashed lines.
The state with $\vec{n}=(2,0,0)$ state is not used in the fit of
the dispersion relation at this $\beta$.}
\label{fig:ep}
\end{figure}

Figure~\ref{fig:ep} shows typical effective mass plots
for pseudoscalar and the vector mesons.
The energies of finite momentum states are successfully
extracted from the region in which the correlator shows plateaus
except the lightest quark region, $\kappa \leq 0.120$, 
which severely suffer from the statistical fluctuation.

\begin{figure}[tb]
\centerline{\psfig{file=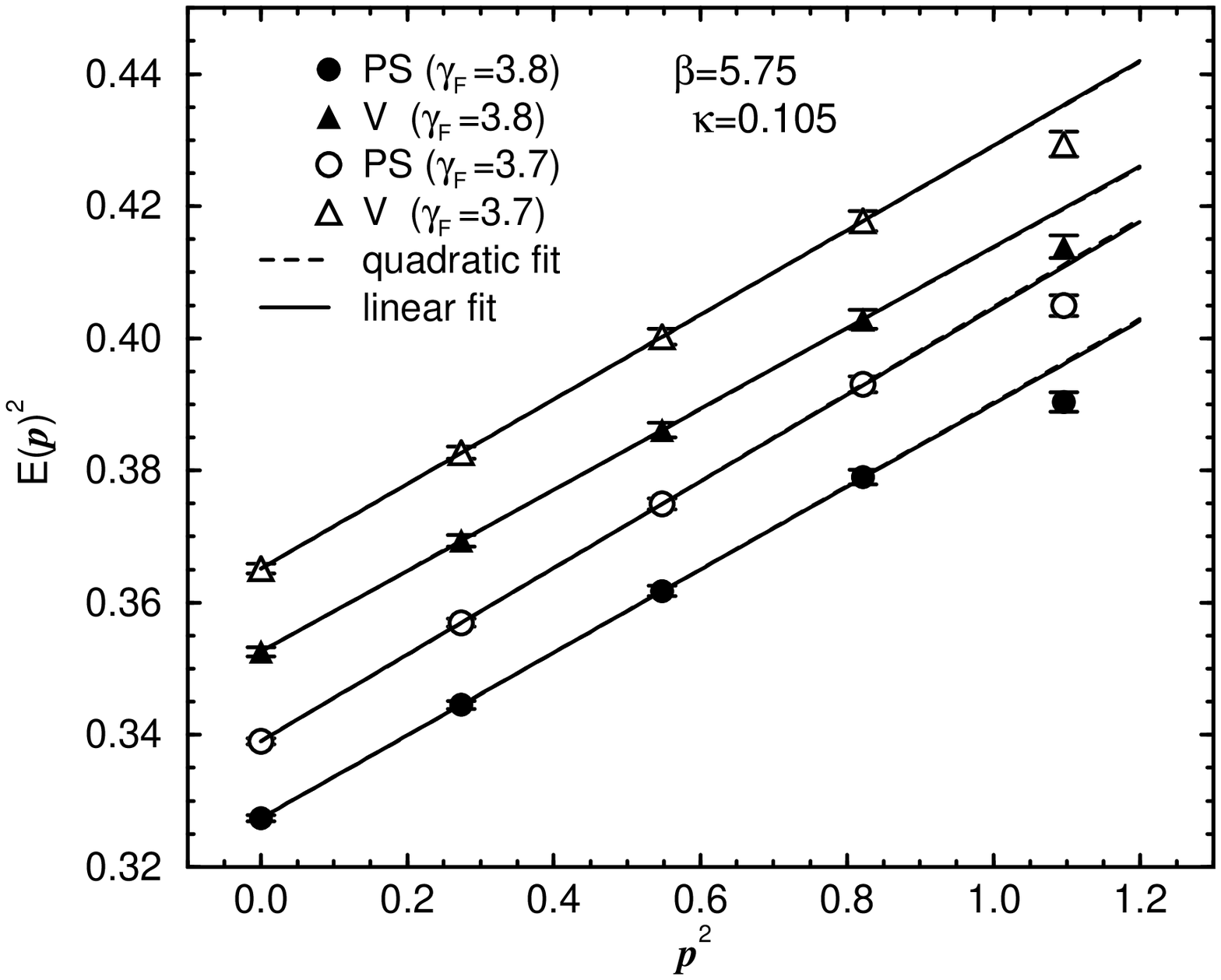,width=0.69\figwidth}
\hspace{-1.2cm}
\psfig{file=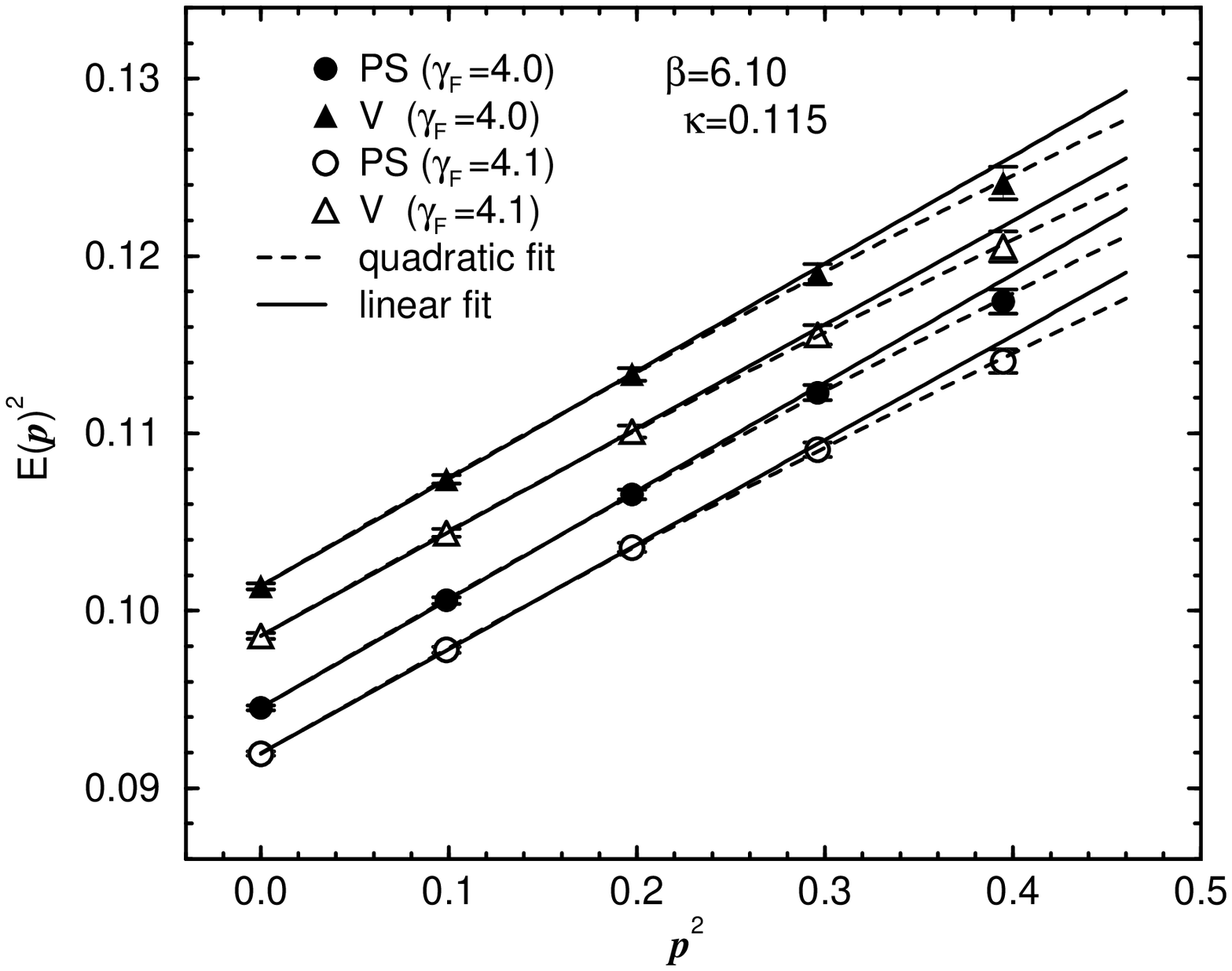,width=0.69\figwidth}}
\vspace{-0.6cm}
\caption{
Dispersion relations of PS and V mesons.
The left figure shows the data at $\kappa=0.105$,
$\gamma_F=3.8$ and $3.7$ on $\beta=5.75$ lattice.
The right figure shows those at $\kappa=0.115$,
$\gamma_F=4.0$ and $4.1$ on $\beta=6.10$ lattice.
Solid lines represent the linear fit and the dashed lines
show the quadratic fit.}
\label{fig:disp}
\end{figure}

The dispersion relation for $\kappa=0.105$ at $\beta=5.75$
is shown in the left figure of Figure~\ref{fig:disp}.
Because of rather large lattice artifact,
the fit to the quadratic form in $\vec{p}^2$ with the energy of
$\vec{n}=(2,0,0)$ state is not a good description of data.
We therefore use only four lowest energy states in
the quadratic fit.
On the other hand, the results of the linear fit (with lowest three
states) and the quadratic fit coincides with good accuracy.
For a few largest hopping parameters, higher momentum states suffers
so large statistical fluctuations that we inevitably adopt the
linear fit.
Since at other values of $\kappa$ the resultant $\xi_F$ coincides
with quadratic fit, we adopt the linear fit for all values
of $\kappa$ at this $\beta$.
Since the same method as $\beta=5.75$ case is adopted at $\beta=5.95$,
we do not repeat the explanation for the fitting procedure.

The right figure of Figure~\ref{fig:disp} shows the dispersion
relation of mesons at $\beta=6.10$ and $\kappa=0.115$,
which corresponds to a similar quark mass as $\kappa=0.105$ at $\beta=5.75$.
The dispersion relation is much improved, and the quadratic fit
is successfully applied including $\vec{n}=(2,0,0)$ state.
Although the difference between the linear fit is small,
as is shown in the figure, we adopt the result of quadratic fit
to determine $\xi_F$ except for most light quark region.
For these three largest $\kappa$,
correlators with $\vec{n}=(2,0,0)$, and occasionally
$(1,1,1)$, suffer from so large statistical fluctuations that
the energy of the states cannot be reliably extracted.
In these cases, we fit the energy to the linear form.

\begin{figure}[tb]
\centerline{\psfig{file=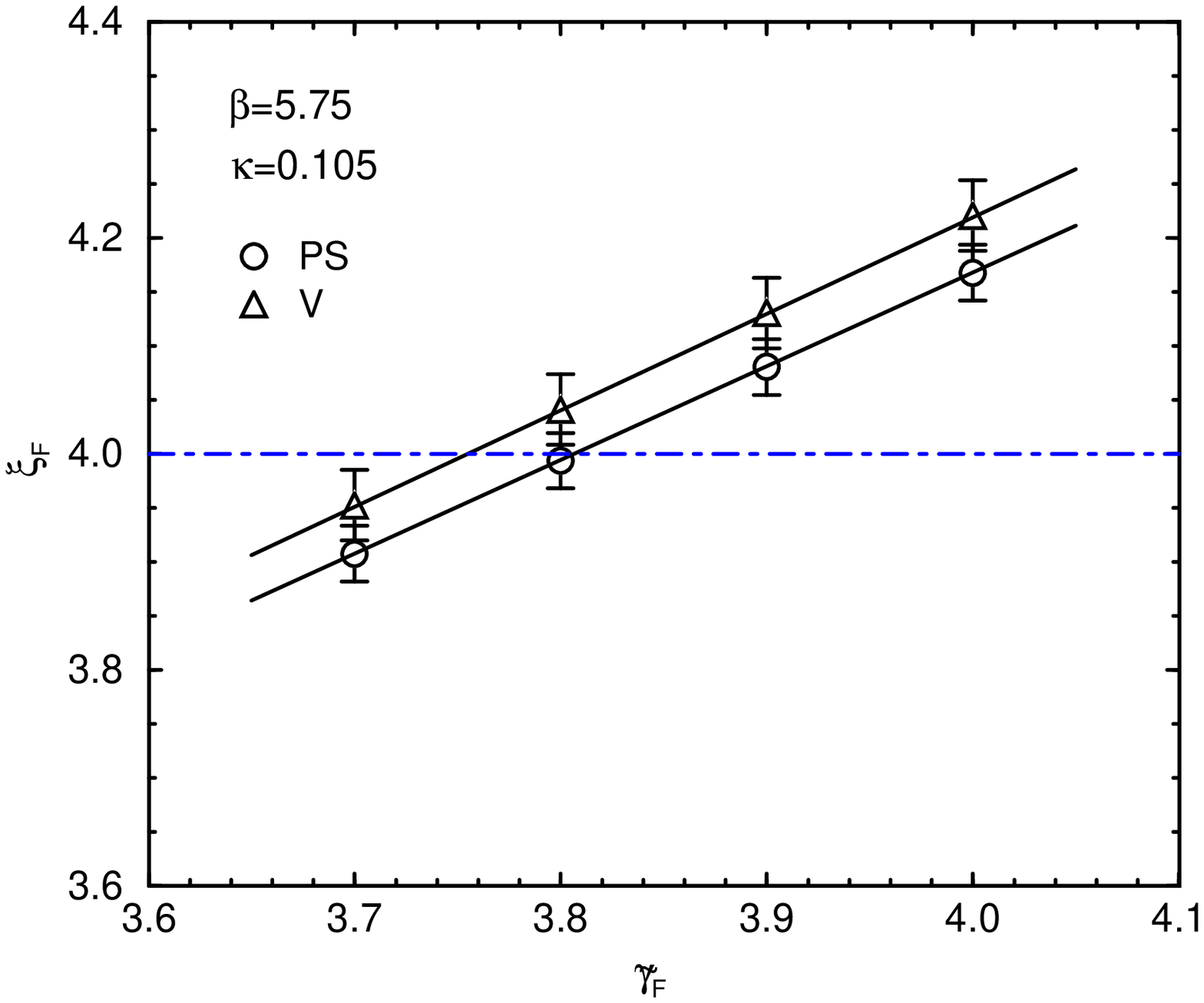,width=0.69\figwidth}
\hspace{-1.2cm}
\psfig{file=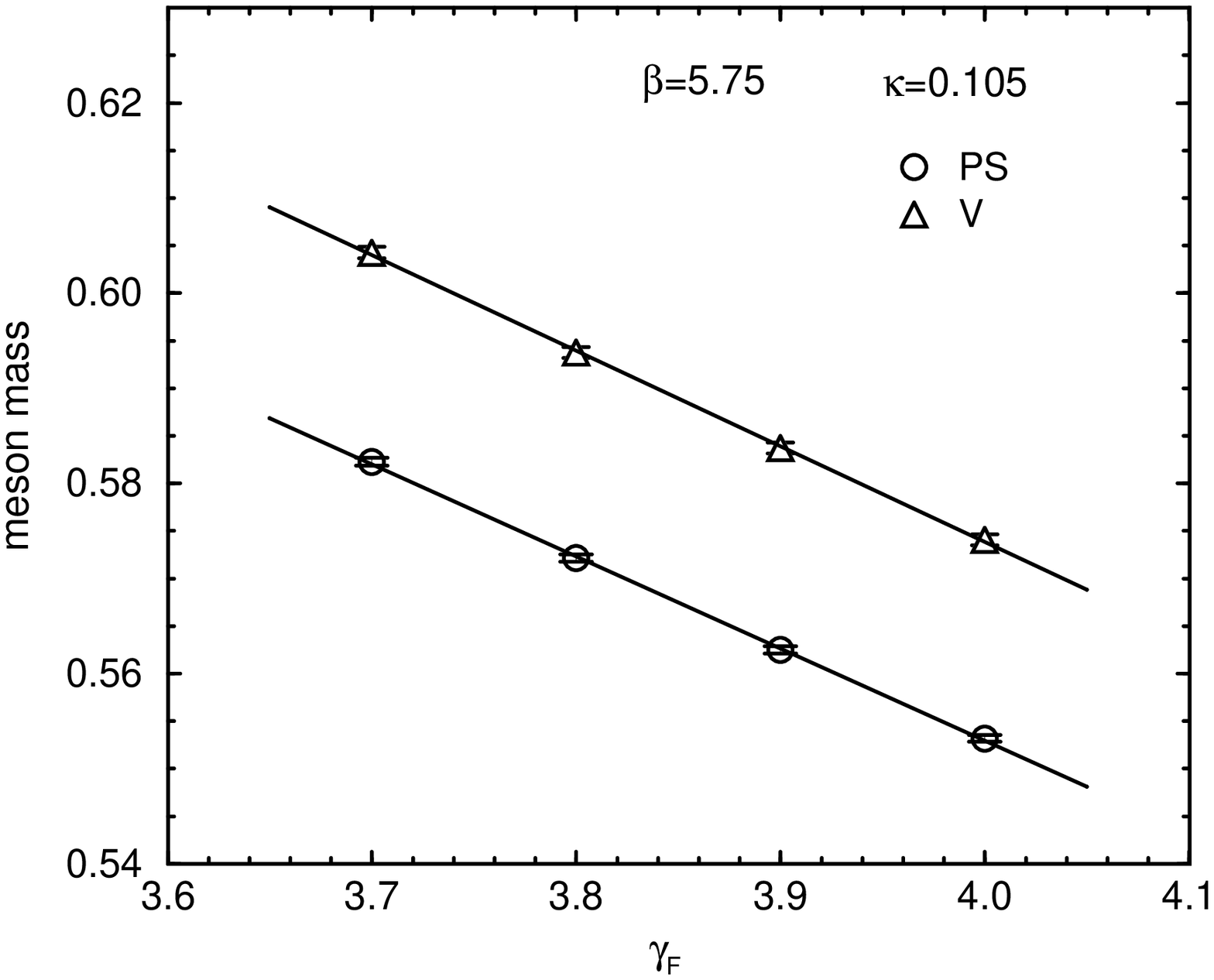,width=0.69\figwidth}}
\vspace{-0.6cm}
\caption{
$\gamma_F$ dependence of $\xi_F$ (left) and meson masses
(right) at $\kappa=0.105$ on $\beta=5.75$ lattice.}
\label{fig:gamma_vs_xi}
\end{figure}

\paragraph{Calibration of $\gamma_F$ for each quark mass.} 

In the left figure of Figure~\ref{fig:gamma_vs_xi},
$\xi_F$ is plotted as the
function of $\gamma_F$ for $\kappa=0.105$ at $\beta=5.75$.
It is clear that $\xi_F$ depends linearly on $\gamma_F$.
The results at $\kappa=0.101$, $0.110$ and $0.112$ also show
similar behavior.
We therefore assume the linear dependence also for other values of
$\kappa$, and interpolate $\xi_F$ to find out $\gamma_F^*$
in each channel.
The numerical results of $\gamma_F^{*(PS)}$, $\gamma_F^{*(V)}$ 
and $\gamma_F^*$ as the average over PS and V mesons are listed
in Table~\ref{tab:qprm_E}--\ref{tab:qprm_G} for each $\beta$.
These tables also show the interpolated masses of PS and V mesons.
We find a tendency that $\gamma_F^{*(PS)}$ is
slightly larger than  $\gamma_F^{*(V)}$ in whole $\kappa$
region. This deviation seems to become smaller for larger $\beta$.
The reason for the discrepancy is understood as the 
systematic errors of $O(\alpha a)$, which will be 
examined in the next subsection in detail.

We also plot the  $\gamma_F$ dependence of meson masses
at $\kappa=0.105$ and $\beta=5.75$
in the right figure of Figure~\ref{fig:gamma_vs_xi}.
This shows that the meson mass is linear in $\gamma_F$ in this range,
and linear interpolation is successfully applicable to determine
the meson masses at $\gamma_F^*$.
Although the $\gamma_F$ dependence of meson mass is in general
unknown for other region of $\kappa$, we expect that the linear
interpolation would work in good accuracy.
The meson masses at $\gamma_F^*$ are also listed
in Table~\ref{tab:qprm_E}--\ref{tab:qprm_G} for each $\beta$.
How the uncertainty in  $\gamma_F^*$
affect on the spectrum is an important problem, 
which will be examined in the next subsection.

\begin{table}[tb]
\caption{
Calibration parameters and result at $\beta=5.75$.
Linear fit is applied to the dispersion relation in the determination
of $\xi_F$.}
\begin{center}
\begin{tabular}{cccccccc}
\hline\hline
$\kappa$  & input $\gamma_F$  & $\!\!N_{conf}\!\!$ &
$\gamma_F^{*(PS)}$ & $\gamma_F^{*(V)}$ &
$\gamma_F^*$  &
$m_{PS}(\gamma_F^*)$ &
$m_{V}(\gamma_F^*)$ \\
\hline
0.124 & 3.9, 4.0 & 400 &
    3.935(77)& 3.83(18) & 3.919(72)& 0.1497( 6)& 0.2294(17) \\
0.122 & 3.9, 4.0 & 400 &
    3.904(48)& 3.884(82)& 3.899(45)& 0.2044( 4)& 0.2650(12) \\
0.120 & 3.9, 4.0 & 400 &
    3.892(43)& 3.888(54)& 3.891(38)& 0.2523( 8)& 0.3018(12) \\
0.118 & 3.9, 4.0 & 400 &
    3.906(36)& 3.894(42)& 3.901(31)& 0.2967( 9)& 0.3387(12) \\
0.116 & 3.9, 4.0 & 300 &
    3.875(35)& 3.841(42)& 3.861(33)& 0.3408(13)& 0.3774(15) \\
0.114 & 3.9, 4.0 & 200 &
    3.899(36)& 3.842(47)& 3.878(36)& 0.3819(17)& 0.4142(19) \\
0.112 & $3.8, 3.9, 4.0$ & 200 &
    3.854(29)& 3.806(37)& 3.836(30)& 0.4252(17)& 0.4546(18) \\
0.110 & $3.8, 3.9, 4.0$ & 200 &
    3.878(33)& 3.827(41)& 3.857(34)& 0.4654(22)& 0.4918(23) \\
0.105 & $3.7, 3.8, 3.9, 4.0$ & 160 &
    3.807(30)& 3.754(37)& 3.786(31)& 0.5738(28)& 0.5954(28) \\
0.101 & $3.7, 3.8, 3.9, 4.0$ & 160 &
    3.730(26)& 3.679(31)& 3.709(27)& 0.6653(30)& 0.6845(30) \\
0.097 & 3.5, 3.6 & 160 &
    3.626(19)& 3.587(24)& 3.611(20)& 0.7647(28)& 0.7823(28) \\
0.095 & 3.5, 3.6 & 160 &
    3.579(18)& 3.540(23)& 3.564(19)& 0.8166(29)& 0.8333(29) \\
0.093 & 3.5, 3.6 & 160 &
    3.530(17)& 3.490(21)& 3.514(18)& 0.8704(29)& 0.8864(29) \\
\hline\hline
\end{tabular}
\end{center}
\label{tab:qprm_E}
\end{table}

\begin{table}[tb]
\caption{
Calibration parameters and result at $\beta=5.95$.
Linear fit is applied to the dispersion relation in the determination
of $\xi_F$.}
\begin{center}
\begin{tabular}{cccccccc}
\hline\hline
$\kappa$  & input $\gamma_F$  & $\!\!N_{conf}\!\!$ &
$\gamma_F^{*(PS)}$ & $\gamma_F^{*(V)}$ &
$\gamma_F^*$  &
$m_{PS}(\gamma_F^*)$ &
$m_{V}(\gamma_F^*)$ \\
\hline
0.124 & 3.9, 4.0 & 500 &
    4.073(95)& 4.15(12) & 4.103(80)& 0.1177( 6)& 0.1649( 9) \\
0.123 & 3.9, 4.0 & 500 &
    4.041(69)& 4.095(93)& 4.060(60)& 0.1456( 2)& 0.1847( 8) \\
0.122 & 3.9, 4.0 & 500 &
    4.029(55)& 4.076(68)& 4.048(48)& 0.1712( 4)& 0.2045( 8) \\
0.120 & 3.9, 4.0 & 500 &
    4.019(36)& 3.996(55)& 4.012(34)& 0.2186( 6)& 0.2444( 9) \\
0.118 & 3.9, 4.0 & 500 &
    4.003(29)& 4.000(35)& 4.002(28)& 0.2625( 8)& 0.2841( 9) \\
0.115 & 3.9, 4.0 & 360 &
    3.992(28)& 3.969(36)& 3.983(29)& 0.3260(12)& 0.3431(13) \\
0.110 & 3.9, 4.0 & 300 &
    3.945(28)& 3.946(36)& 3.946(29)& 0.4297(19)& 0.4427(19) \\
0.107 & 3.9, 4.0 & 300 &
    3.910(25)& 3.911(31)& 3.910(26)& 0.4930(20)& 0.5046(21) \\
0.104 & 3.9, 4.0 & 200 &
    3.876(28)& 3.875(36)& 3.876(30)& 0.5573(27)& 0.5677(27) \\
0.102 & 3.9, 4.0 & 200 &
    3.848(26)& 3.847(34)& 3.847(28)& 0.6016(28)& 0.6113(28) \\
0.100 & 3.9, 4.0 & 200 &
    3.815(25)& 3.816(32)& 3.815(27)& 0.6470(29)& 0.6562(29) \\
0.097 & 3.8, 3.9 & 200 &
    3.766(24)& 3.765(30)& 3.766(26)& 0.7178(32)& 0.7264(32) \\
0.093 & 3.7, 3.8 & 200 &
    3.688(23)& 3.687(29)& 3.688(25)& 0.8180(37)& 0.8257(37) \\
\hline\hline
\end{tabular}
\end{center}
\label{tab:qprm_F}
\end{table}

\begin{table}[tb]
\caption{
Calibration parameters and result at $\beta=6.10$.
For $\kappa=$0.124, 0.123 and 0.122, the dispersion relation is fitted
to the linear form in determining $\xi_F$.
For the remaining $\kappa$'s, quadratic fit is applied.}
\begin{center}
\begin{tabular}{cccccccc}
\hline\hline
$\kappa$  & input $\gamma_F$  & $\!\!N_{conf}\!\!$ &
$\gamma_F^{*(PS)}$ & $\gamma_F^{*(V)}$ &
$\gamma_F^*$  &
$m_{PS}(\gamma_F^*)$ &
$m_{V}(\gamma_F^*)$ \\
\hline
0.124& 4.0, 4.1& 600&
    4.020(77)& 3.63(27) & 3.991(79)& 0.1008( 5)& 0.1379( 7) \\
0.123& 4.0, 4.1& 600&
    4.005(52)& 3.860(98)& 3.973(55)& 0.1294( 1)& 0.1582( 5) \\
0.122& 4.0, 4.1& 600&
    3.998(41)& 3.913(69)& 3.976(44)& 0.1549( 3)& 0.1787( 6) \\
0.120& 4.0, 4.1& 400&
    4.040(29)& 4.064(45)& 4.047(30)& 0.2007( 6)& 0.2182( 7) \\
0.118& 4.0, 4.1& 200&
    4.040(33)& 4.014(52)& 4.032(34)& 0.2440( 9)& 0.2579(10) \\
0.115& 4.0, 4.1& 200&
    4.024(28)& 4.008(41)& 4.019(30)& 0.3067(12)& 0.3176(13) \\
0.110& 4.0, 4.1& 200&
    4.013(38)& 3.996(54)& 4.007(43)& 0.4078(26)& 0.4160(27) \\
0.107& 4.0, 4.1& 200&
    3.988(36)& 3.986(46)& 3.987(39)& 0.4694(29)& 0.4766(30) \\
0.104& 4.0, 4.1& 200&
    3.951(33)& 3.955(42)& 3.952(36)& 0.5331(31)& 0.5396(31) \\
0.102& 3.9, 4.0& 200&
    3.918(32)& 3.928(40)& 3.922(35)& 0.5773(34)& 0.5834(34) \\
0.100& 3.9, 4.0& 200&
    3.877(25)& 3.883(32)& 3.879(27)& 0.6240(28)& 0.6298(28) \\
0.097& 3.9, 4.0& 200&
    3.834(22)& 3.839(28)& 3.836(24)& 0.6931(28)& 0.6984(28) \\
0.093& 3.8, 3.9& 200&
    3.769(21)& 3.776(26)& 3.772(23)& 0.7903(32)& 0.7953(31) \\
\hline\hline
\end{tabular}
\end{center}
\label{tab:qprm_G}
\end{table}

\begin{figure}[tb]
\centerline{\psfig{file=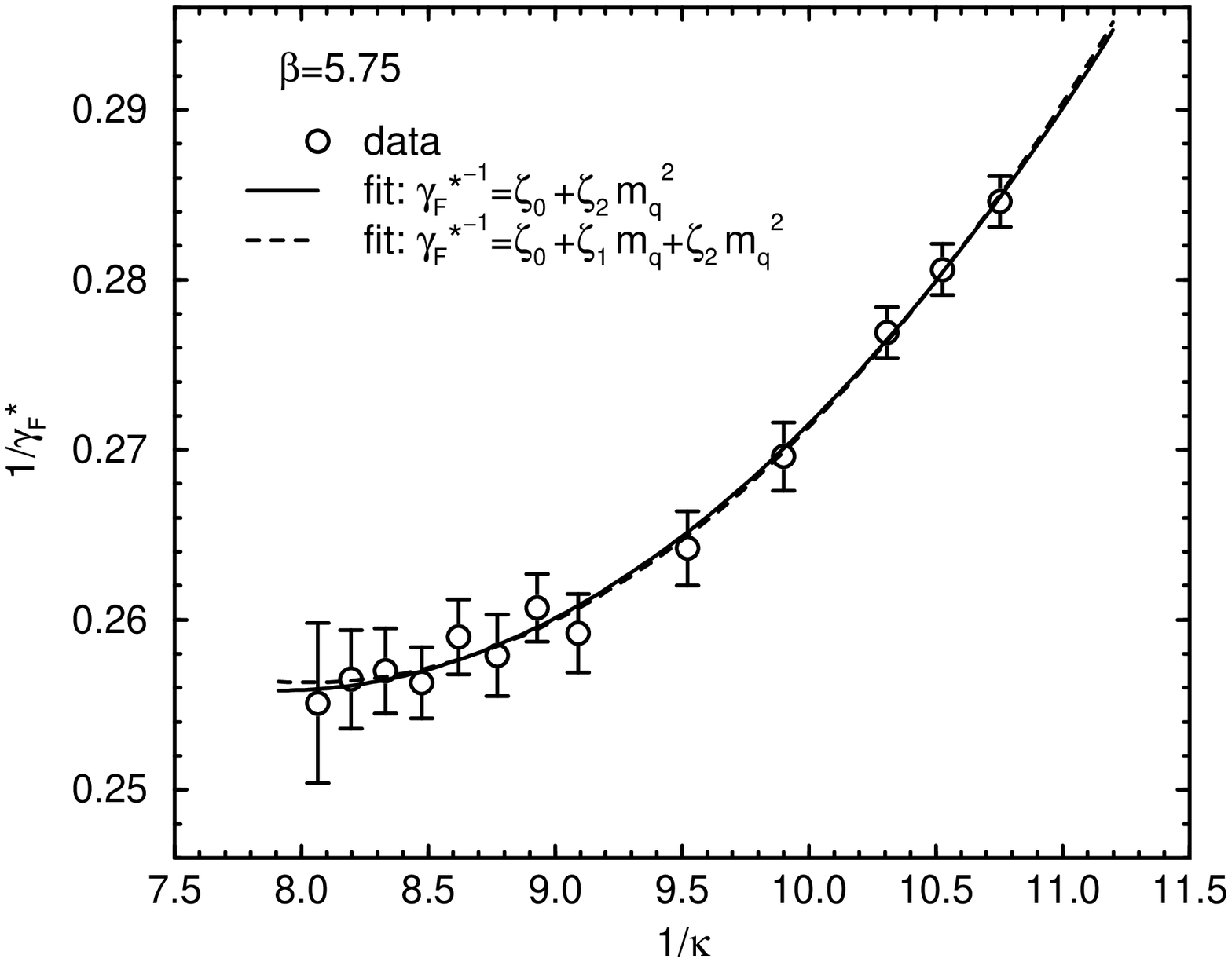,width=0.75\figwidth}}
\vspace{-1cm}
\centerline{\psfig{file=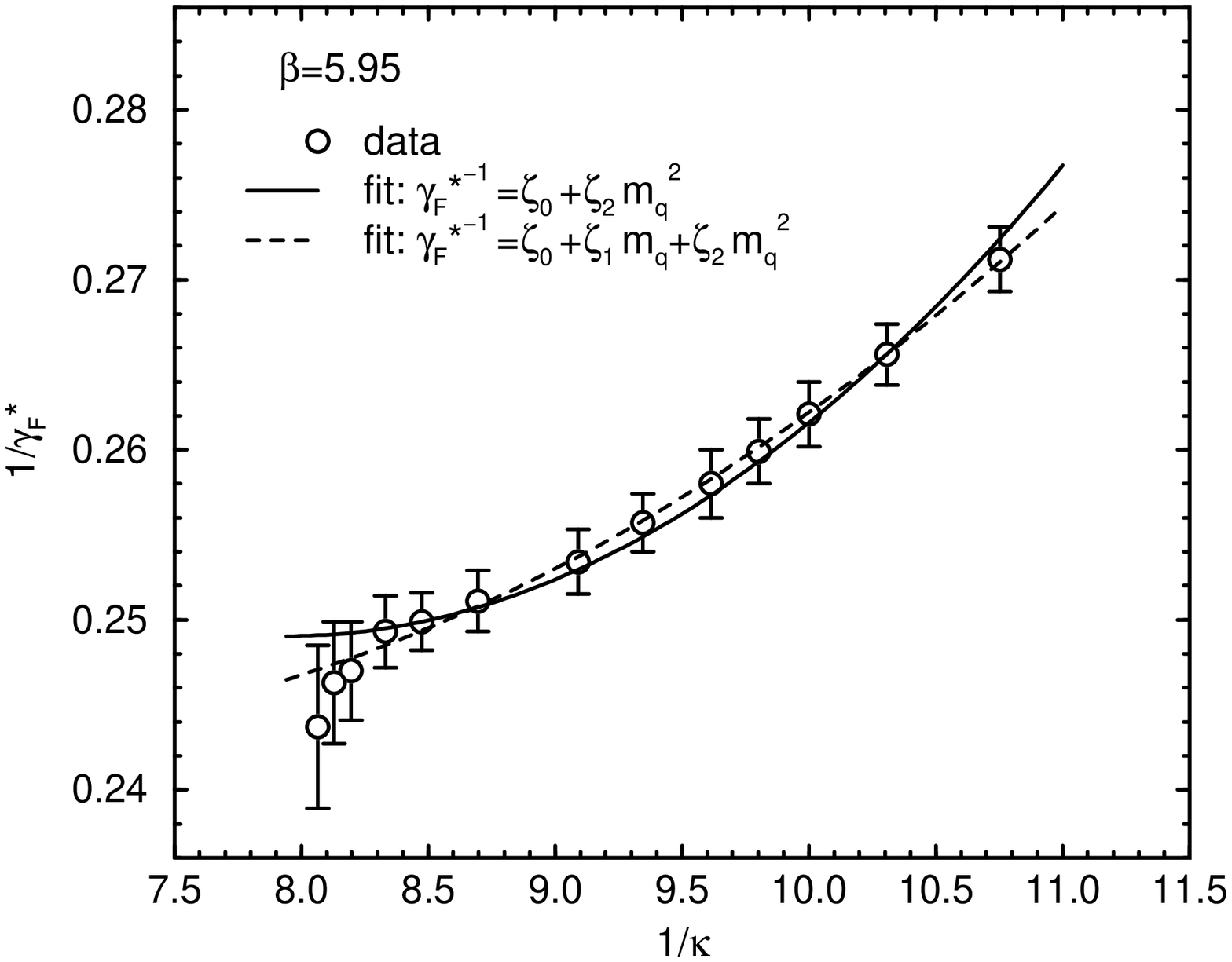,width=0.75\figwidth}}
\vspace{-1cm}
\centerline{\psfig{file=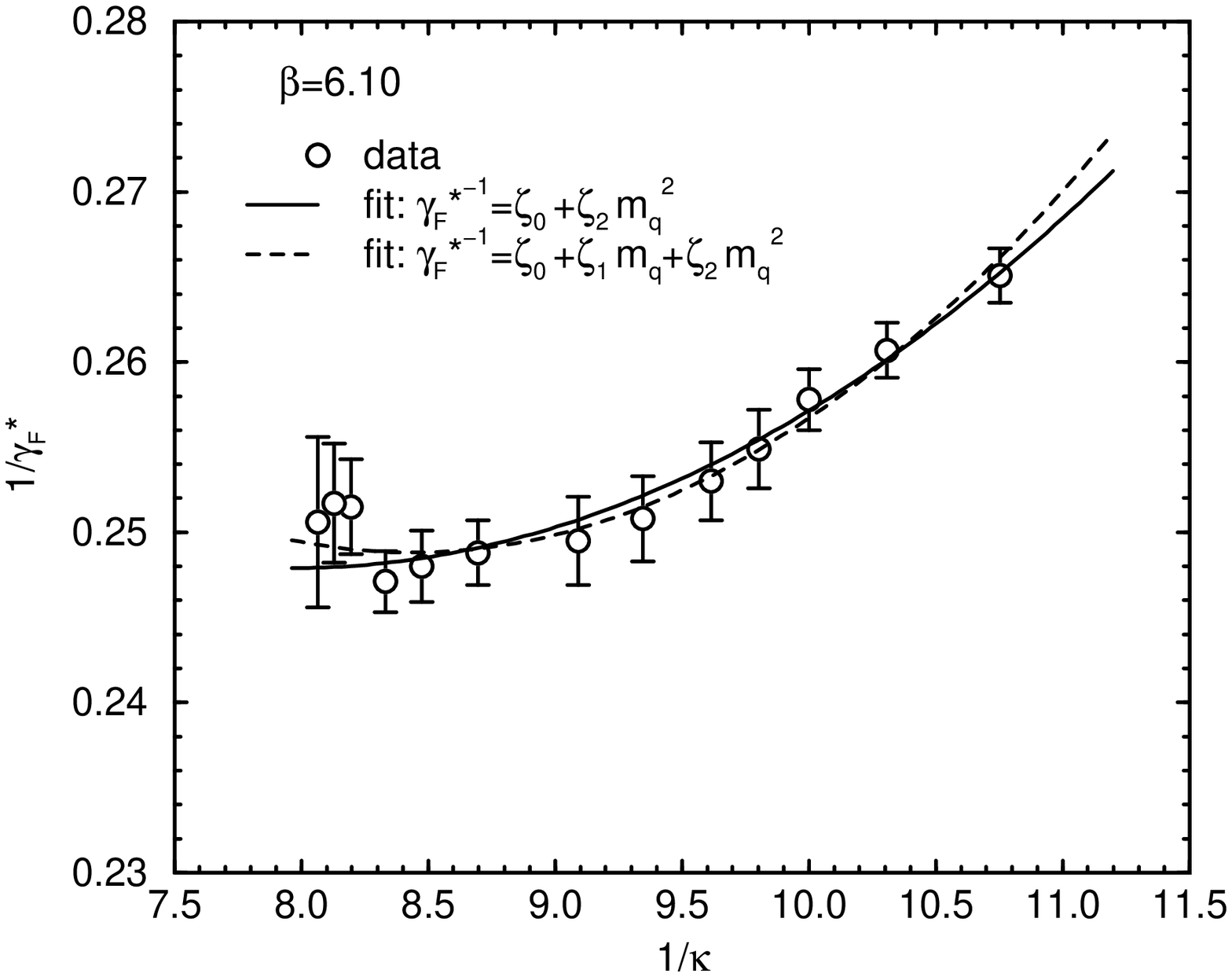,width=0.75\figwidth}}
\vspace{-0.6cm}
\caption{$1/\gamma_F^*$ vs $1/\kappa$ at each $\beta$.
Solid lines show the fit linear in $m_q^2$
while the dashed lines represent the fit quadratic in $m_q$.}
\label{fig:zeta}
\end{figure}

\paragraph{Fit of $\gamma_F^*$.} 

To represent  $\gamma_F^*$ as the function of $\kappa$,
we introduce the quark mass as
\begin{equation}
 m_q = \frac{1}{2\xi}
   \left( \frac{1}{\kappa}-\frac{1}{\kappa_c} \right).
\end{equation}
This is similar relation as $m_0$, the bare quark mass in the temporal
lattice unit, while the present form is independent of
$\gamma_F$.
$\kappa_c$ is determined from massless point of the pseudoscalar
meson mass.
We extrapolate $m_{PS}^2$ linearly in $1/\kappa$ using two
largest values of $\kappa$, and find as $\kappa_c=$
0.12640(5) at $\beta=5.75$,  0.12592(6) at $\beta=5.95$ and 
0.12558(4) at $\beta=6.10$.

In the calibration at each $\kappa$, we found that
the value of $\gamma_F^*$ is easily determined 
precisely (to the level of 1 \%), while it become difficult
with increasing $\kappa$ toward $\kappa_c$.
However, it is expected $\gamma_F^*$
smoothly approaches to certain definite value in the limit of 
$m_q\rightarrow 0$,
since in this limit our form of action is simply
a direct generalization of clover quark action on the anisotropic lattice.
In fact, as shown in Subsection~\ref{subsec:free_quark},
$\gamma_F^*$ linearly depends on $m_0^2$ at the tree level.
In taking the limit of $m_q\rightarrow 0$ for $\gamma_F^*$,
the precise values of mean-filed values do not matter,
since in the definition of $\gamma_F$, tadpole-improvement is
applied as a multiplicative factor,
$\eta_{MF}=u_{\tau}/u_{\sigma}$.
Therefore
the most reliable way to determine the value of $\gamma_F^*$
in the light quark region is a global fit of  $\gamma_F^*$,
assuming appropriate form of $m_q$
dependence.

We fit the result of  $1/\gamma_F^*$ to the linear form in $m_q^2$
and the quadratic form in $m_q$.
The result of fit is listed in Table~\ref{tab:fit_gamma_F}
and also shown in Figures~\ref{fig:zeta} as the solid
and the dashed curves.
Since the obtained points for $\gamma_F^*$ is with different
number of configurations, these points correlate in not obvious way.
We quote the errors and $\chi^2$ of uncorrelated fit in
Table~\ref{tab:fit_gamma_F}.
As shown in these table and figures, the linear fit in $m_q^2$
well represents the data.
The values of $\zeta_0$, which is the $1/\gamma_F^*$ in the chiral limit, 
is close to the tree level result $1/\xi = 1/4$, which
implies that the tadpole-improvement works well.
Apart from the lightest quark mass region, $\gamma_F^*$ is determined
within 1 \% accuracy, and there the curves of two fits are consistent
with each other.
In approaching the chiral limit, there is a systematic error
concerning the fit form, as well as the statistical error.
We estimate the latter by the error of fit in $\zeta_0$ (from
the quadratic fit in $m_q^2$), and as about 1 \%.
This relatively small statistical error is due to the global fit of
$1/\gamma_F^*$ with assumed form of $m_q$ dependence.
The systematic error in adopting specific form of fit is estimated
by the difference between these two fits, and also results in
1 \% level.
Adopting the linear form in $m_q^2$, we conclude that
the $\gamma_F^*$ is determined under the assumed dispersion relation
within 1 \% statistical accuracy in the whole quark
mass region less than around the charm quark mass,
while in the chiral limit there is additional  1 \% systematic
uncertainty concerning the form of fit.

\begin{table}[tb]
\caption{
Fit results of $\gamma_F$.}
\begin{center}
\begin{tabular}{ccccccc}
\hline\hline
$\beta$ &  fit type &
$\zeta_0$ & $\zeta_1$ & $\zeta_2$ &
$\chi^2 / N_{df}$  & $\gamma_F^*(m_q=0)$  \\
\hline
5.75& linear& 0.2558( 9)&       --    & 0.230(12)& 1.83 / 11& 3.909(14)\\
    & quad. & 0.2564(23)& $-$0.007(28)& 0.247(68)& 1.77 / 10& 3.901(34)\\
\hline
5.95& linear& 0.2490( 8)&       --    & 0.189(15)& 3.52 / 11& 4.016(13)\\
    & quad. & 0.2465(18)& \  0.036(23)& 0.095(61)& 1.01 / 10& 4.057(30)\\
\hline
6.10& linear& 0.2479( 9)&     --      & 0.143(14)& 4.44 / 11& 4.034(14)\\
    & quad. & 0.2493(18)& $-$0.022(24)& 0.200(63)& 3.55 / 10& 4.011(28)\\
\hline\hline
\end{tabular}
\end{center}
\label{tab:fit_gamma_F}
\end{table}

\subsection{Uncertainties in calibration}

In the last subsection, we determined $\gamma_F^*$
as a global function of $m_q$.
This expression inevitably suffer from systematic uncertainties
as well as the statistical uncertainty:
\begin{equation}
 \gamma_F^*   =  \gamma_F^{*(prop)}
              + \delta \gamma_F^{(stat)}
              + \delta \gamma_F^{(O(\alpha a))}
              + \delta \gamma_F^{(O(a^2))} 
             (  + \delta \gamma_F^{(chiral)} ) .
\end{equation}
$\gamma_F^{*(prop)}$ represent the proper value of the bare anisotropy.
$\delta \gamma_F^{(stat)}$ is the statistical error in determination 
of $\gamma_F^*$, and at 1 \% level.
The last two terms are main sources of systematic errors
due to finite lattice artifact.
The first one, $\delta \gamma_F^{(O(\alpha a))}$, is from the
tree-level approximation of clover coefficients.
We estimate the size of this error by the difference between the
values of $\gamma_F^*$ determined with PS and V mesons.
The second systematic error, $\delta \gamma_F^{(O(a^2))}$,
is estimated by comparing the results of calibration
from two different forms of dispersion relations which differ by $O(a^2)$.
In addition to these systematic uncertainties, in the chiral limit
there is also a systematic error concerning the form of fit
of $\gamma_F^*$ in $m_q$.

Another important subject is to estimate
how the observables are affected by the uncertainty in $\gamma_F^*$.
We study the response of meson masses with respect to the change of
$\gamma_F^*$ at each $\kappa$, from which 
the effect of $\gamma_F$ on meson masses for given quark mass is
approximately estimated.
Strictly speaking, changing $\gamma_F$ for a fixed $\kappa$ induces
a slight change in the quark mass, hence the above analysis is
adequate for relatively heavier quark mass region, such as
$m_s < m_q$.
We postpone the study of the effect on the light hadron
spectrum to the end of next section.

\paragraph{Difference between $\gamma_F^{*(PS)}$ and
            $\gamma_F^{*(V)}$.}

\begin{figure}[tb]
\centerline{\psfig{file=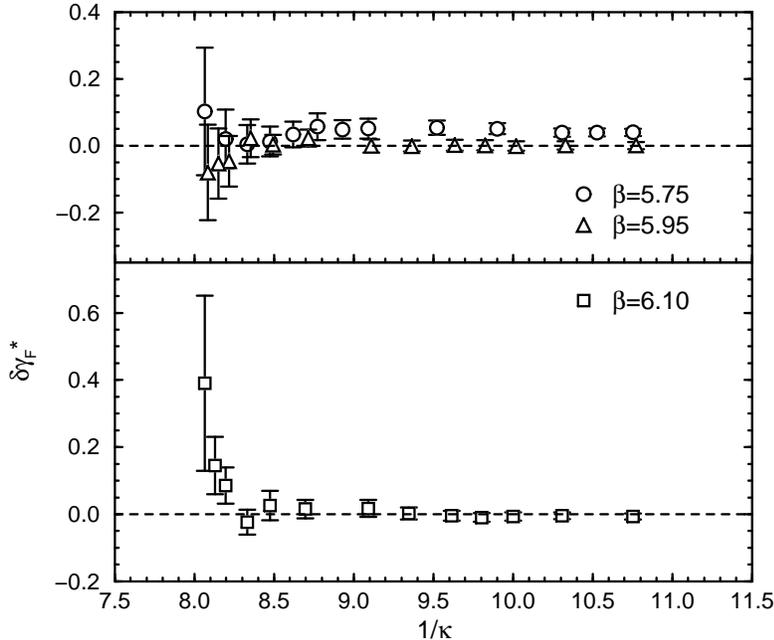,width=0.9\figwidth}}
\vspace{-0.6cm}
\caption{$\delta\gamma_F^*=\gamma_F^{*(PS)}
 - \gamma_F^{*(V)}$ at $\beta=$5.75 and 5.95
in upper part, and 6.10 in lower part.}
\label{fig:dgamma}
\end{figure}

Since we use the $O(a)$ improved quark action, the main contribution
from the $O(a)$ lattice artifact is absent.
However, since the clover coefficients are not tuned beyond the
tree-level, $O(\alpha a)$ error still remains, although
the tadpole-improvement would partially removes this effect.
An appropriate probe of this systematic effect on the calibration
is the difference between $\gamma_F^*$'s of the pseudoscalar
and the vector mesons.
Figure~\ref{fig:dgamma} shows
$\delta\gamma_F^* \equiv
\gamma_F^{*(PS)} - \gamma_F^{*(V)}$.
At $\beta=5.75$, there is a systematic difference of
$\delta\gamma_F^*$ from zero except the small quark mass
region, where the statistical error is dominant.
At $\beta=5.95$, $\delta\gamma_F^*$ is consistent with zero
in whole $\kappa$ region.
This implies that the $O(\alpha a)$ error in the calibration is
sufficiently reduced at this $\beta$.
In the case of $\beta=6.10$,  $\delta\gamma_F^*$ is
also consistent with zero except the lightest quark region.
In this region, precise determination of energy at finite
momentum states is difficult due to the statistical fluctuation and
hence the resultant $\gamma_F$ contains large uncertainty.
We regard that the $O(\alpha a)$ effect in the calibration is
also small at this $\beta$.

\paragraph{$O(a^2)$ systematic uncertainty.}

\begin{figure}[tb]
\centerline{\psfig{file=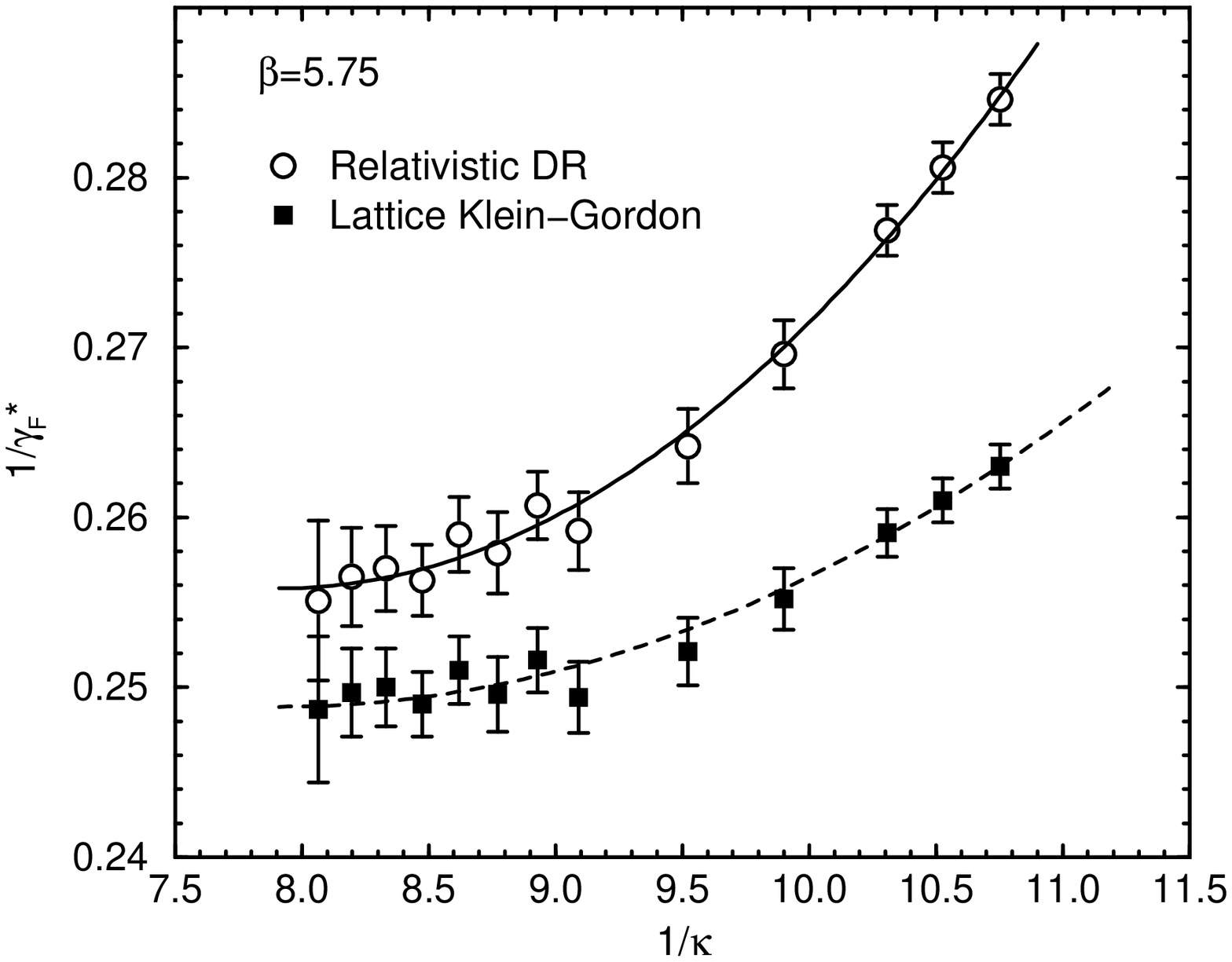,width=0.69\figwidth}
\hspace{-1.2cm}
\psfig{file=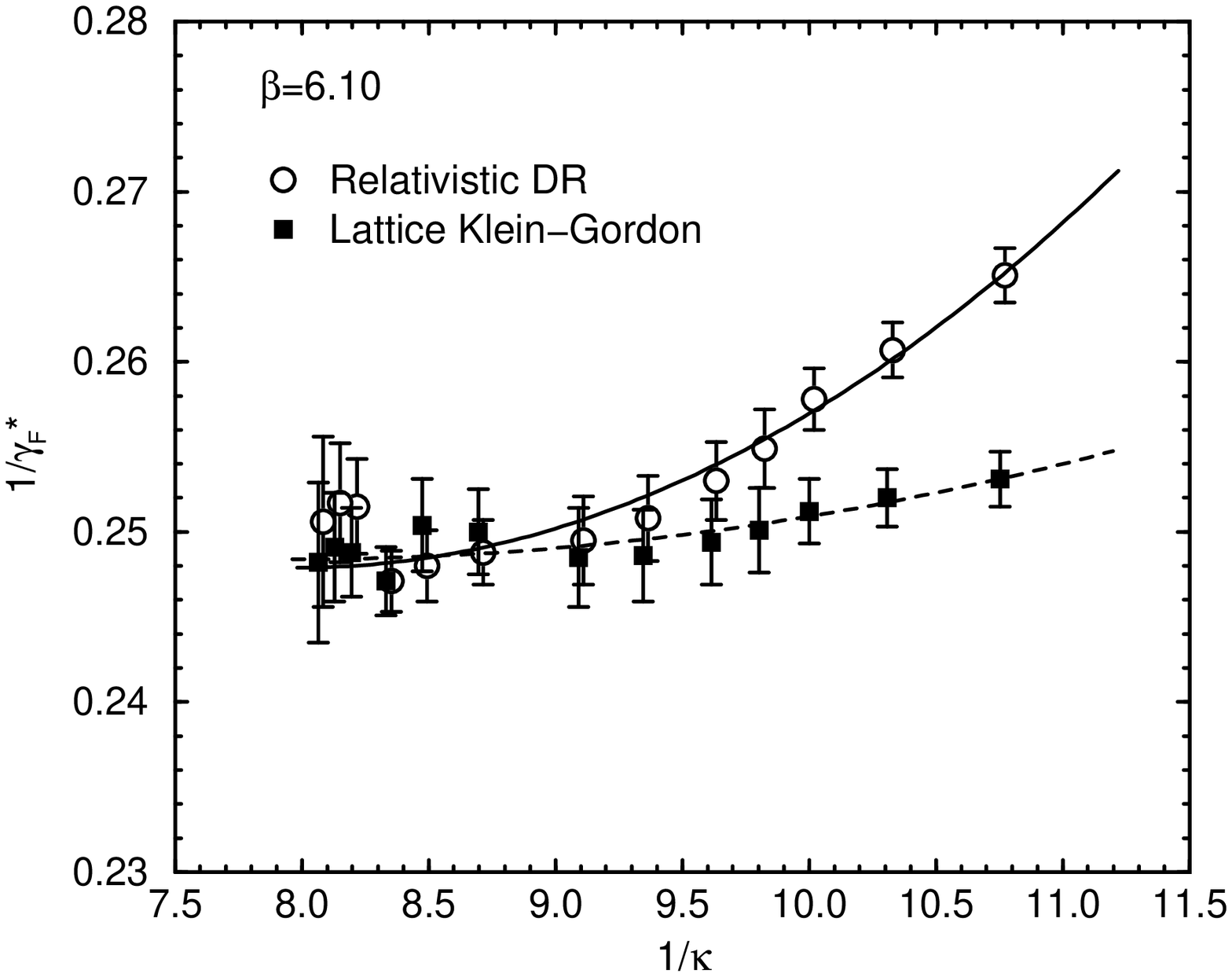,width=0.69\figwidth}}
\vspace{-0.6cm}
\caption{
Comparison of the results of calibrations with two types
of dispersion relations.
The curves represent the the result of linear fit in $m_0^2$.
Solid lines are the fit results with relativistic dispersion relation 
and the dashed lines are those with Lattice Klein-Gordon dispersion 
relation.}
\label{fig:KGDR}
\end{figure}

Although we employed the continuum dispersion relation,
this introduces a systematic error
of $O(a^2)$ to the calibration of $\gamma_F$.
In order to estimate the typical size of this error, we compare
$\gamma_F^*$ determined above with $\gamma_{F(KG)}^*$, the result
obtained using the dispersion relation from the
lattice Klein-Gordon action, ($\ref{eq:DR2}$).
Figure~\ref{fig:KGDR} shows this comparison
at $\beta=5.75$ and $6.10$.
In extracting $\xi_{KG}$ from the Klein-Gordon dispersion relation,
we fit $\cosh E(\vec{p})$ to the linear form in
$\sum_i \sin^2(p_i/2)$ using lowest three momentum states.
As explained in Section~\ref{sec:calibration},
expected difference of $\xi_F$ and $\xi_{KG}$ is $O(m^2)$,
where $m$ is the meson mass.
Although the explicit relation between $\gamma_{F(KG)}^*$ and 
$\gamma_F^*$ is unknown, one can expect that the difference of
them is also in $O(m^2)$, and hence increase with increasing
quark mass.
This behavior is clearly observed in Figure~\ref{fig:KGDR}.
Table~\ref{tab:fit_gamma_F_KG} is the result of fit of
$\gamma_{F(KG)}^*$ to the linear form in $m_q^2$.

We find small difference between the results with relativistic and
Klein-Gordon dispersion relations in the small quark mass region.
This difference decreases as increasing $\beta$, and seems to
be sufficiently reduced at $\beta=6.10$.
The typical size of difference in $\gamma_F^*$'s at the chiral limit is
less than 3, 2, 1 \% at $\beta=5.75$, 5.95 and 6.10, respectively.
The important feature is that two procedures tend to coincide
with each other with increasing $\beta$.
We also observe that the difference between $\gamma_F^*$ and
$\gamma_{F(KG)}^*$ 
increases in the large quark mass region, $m_q > 0.2 a_{\tau}^{-1}$.
This is consistent behavior, since there the Klein-Gordon dispersion
relation fails to incorporate the quark mass dependence properly,
and $\gamma_{F(KG)}^*$ is expected to be larger than
$\gamma_F^*$ in $O(m^2)$.
Therefore we conclude that the uncertainty due to the assumed form
of the meson dispersion relation is under control and smoothly
disappears in approaching the continuum limit.

\begin{table}[tb]
\caption{
The result of linear fit in $m_q^2$ of $\gamma_{F(KG)}^*$,
the tuned bare anisotropy with the Klein-Gordon dispersion relation.}
\begin{center}
\begin{tabular}{cccccc}
\hline\hline
$\beta$ &  fit type &
$\zeta_0$ & $\zeta_2$ &
$\chi^2 / N_{df}$  & $\gamma_F^*(m_q=0)$  \\
\hline
5.75& linear& 0.2488( 8)& 0.112(11)& 2.33 / 11& 4.019(13) \\
\hline
5.95& linear& 0.2446( 8)& 0.071(14)& 2.14 / 11& 4.088(13)\\
\hline
6.10& linear& 0.2484(10)& 0.039(15)& 1.59 / 11& 4.026(16)\\
\hline\hline
\end{tabular}
\end{center}
\label{tab:fit_gamma_F_KG}
\end{table}

\paragraph{Uncertainty in meson mass due to calibration error.}

Another important issue on the systematic errors is how the
uncertainty in $\gamma_F$ is transmitted to the observables.
As an important example, here we focus on the effect on the
meson masses. 

\begin{figure}[tb]
\centerline{\psfig{file=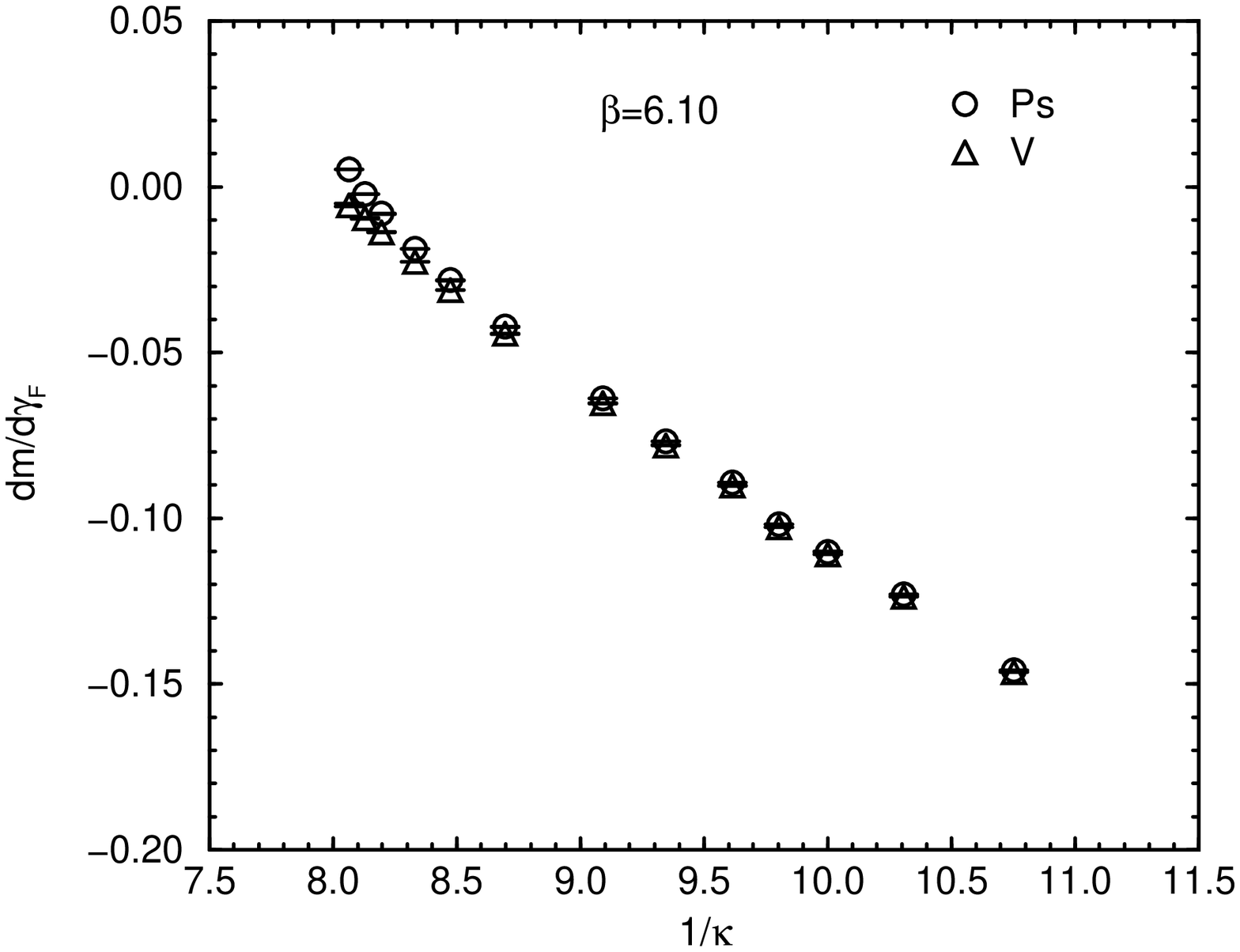,width=0.65\figwidth}
\hspace{-0.8cm}
\psfig{file=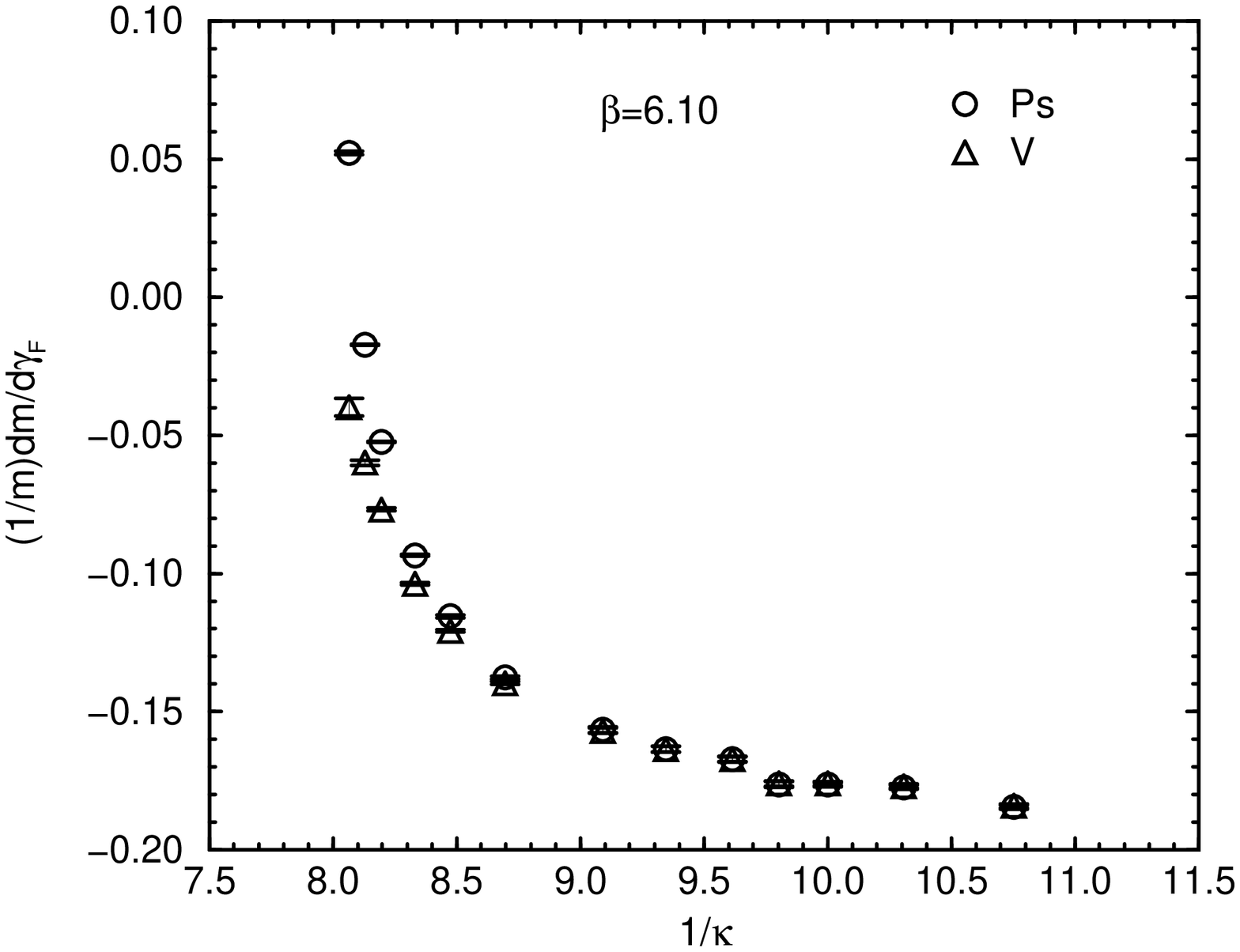,width=0.65\figwidth}}
\vspace{-0.6cm}
\caption{
The response of meson masses to the change of $\gamma_F$
at $\beta=6.10$.
The left figure shows $dm/d\gamma_F$, while the ratio 
$(1/m) dm/d\gamma_F$ is shown in the right figure.
the result for the pseudoscalar and the vector mesons 
are represented by circles and triangles respectively.}
\label{fig:dmdgamma}
\end{figure}

Since we linearly interpolate the meson masses in $\gamma_F$,
we obtain $d m/ d \gamma_F$ at 
$\gamma_F=\gamma_F^*$ from the slope of the linear fit.
In Figure~\ref{fig:dmdgamma}, we show
$d m/ d \gamma_F$, the response of meson mass
to the bare anisotropy, at $\beta=6.10$ in two ways.
Similar feature is found in the results at $\beta=5.75$ and
$\beta=5.95$.
In the left figure $d m/ d \gamma_F$ is
shown as the function of $1/\kappa$.
In the case of vector meson, it seems to decrease linearly
in quark mass from zero at the massless limit.
On the other hand, for the pseudoscalar meson, 
$d m/ d \gamma_F$ slightly positive 
in the vicinity of $1/\kappa_c$.
This behavior may be due to the uncertainty in the definition
of $\kappa$, because if $\kappa$ is properly related to the fixed
quark mass, increasing $\gamma_F$ implies to increase
the propagation in temporal direction, hence it corresponds to
decreasing quark mass.
Therefore, the present analysis may not be adequate for estimating
the response of masses with respect to $\gamma_F$
in the vicinity of the chiral limit.
Observing Figure~\ref{fig:dmdgamma}, one can find that the range of
quark mass larger than the strange quark mass do not suffer from
the ambiguity in the definition of $\kappa$.

We have no clear explanation for that
$d m/ d \gamma_F$ seems to be proportional to
the quark mass.
Practically,
it is a good feature that the ambiguity of $\gamma_F^*$
has only little effects on the meson mass in the small but nonzero 
quark mass region,
since there relative change of mass is significant.
Except the lightest quark mass region, the determination of
$\gamma_F^*$ is directly performed with 1 \% accuracy,
which means the uncertainty of $\gamma_F^*$ is around 0.04.
The right figure in Figure~\ref{fig:dmdgamma} implies that
the uncertainty in the meson mass is less than 1 \%.
This uncertainty is most severe in the heavy quark region,
and become milder as quark mass decreases.

While we find that the meson masses at certain $\kappa$ is 
not so sensitive to the uncertainty of $\gamma_F^*$, 
the same argument does not hold for the chiral limit.
Since the pseudoscalar meson mass becomes zero in the chiral limit, 
the relative uncertainties $\delta m_{PS}/m_{PS}$ for a fixed 
$\kappa$ near the chiral limit is of course very large. 
However, this is not the correct way to estimate the uncertainties 
in the mass spectrum in the chiral limit. What one is really interested 
in for the chiral limit is not the change the hadron masses 
including the pion mass for a fixed $\kappa$ but the change of hadron 
masses except the pion mass at the point where the pion becomes massless.
Since the critical hopping parameter $\kappa_c$ is affected by
the change in $\gamma_F$, one need to treat the chiral limit carefully.
In the next section, we discuss the uncertainties of the 
hadron spectrum in the chiral limit based on the extrapolation
in terms of the pseudoscalar meson mass squared
instead of  $1/\kappa$.

\subsection{Summary of calibration}
\label{subsec:calbration_summary}

In this section, we have implemented the anisotropic $O(a)$ improved
Wilson action in the region of quark mass up to around the charm
quark mass, at three values of $\beta$ at $\xi=4.0$.
The fermionic anisotropy $\xi_F$ is extracted from the meson dispersion
relation.
Then we find the value of bare anisotropy parameter, 
$\gamma_F^*$, at which $\xi_F=\xi$ holds.
The value of $\gamma_F^*$ in the massless limit is obtained by
extrapolating the data by fitting to the linear form in $m_q^2$,
where $m_q$ is naively defined quark mass.
This is the most reliable way to determine $\gamma_F^*$
for small quark mass region, since there the statistical fluctuation
in finite momentum states is severely large.
The fit of $1/\gamma_F^*$ to the linear form in $m_q^2$ seems quite
successful, and $\gamma_F^*$ at the chiral limit is close to the tree-level
value, $\xi$.
The statistical uncertainty in $\gamma_F^*$ is estimated as
in the order of 1 \% in whole explored quark mass region.
In the chiral limit, there is also 1 \% systematic uncertainty
concerning the form of fit.
Here we summarize the main result of calibration, the expression of
$\gamma_F^*$ for a given $\kappa$:
\begin{equation}
 \frac{1}{\gamma_F^*}(m_q) = \zeta_0 + \zeta_2 m_q^2,
  \hspace{1cm}
 m_q = \frac{1}{2 \xi} \left( \frac{1}{\kappa} 
    - \frac{1}{\kappa_c} \right),
\end{equation}
\begin{eqnarray}
 \beta=5.75: & & \zeta_0 = 0.2558(9), \hspace{0.5cm}
  \zeta_2 = 0.230(12), \hspace{0.5cm} \kappa_c = 0.12640(5) \\
 \beta=5.95: & & \zeta_0 = 0.2490(8), \hspace{0.5cm}
  \zeta_2 = 0.189(15), \hspace{0.5cm} \kappa_c = 0.12592(6) \\
 \beta=6.10: & & \zeta_0 = 0.2479(9), \hspace{0.5cm}
  \zeta_2 = 0.143(14), \hspace{0.5cm} \kappa_c = 0.12558(4).
\end{eqnarray}

To examine the uncertainty in the calibration, we have also carried
out the following analysis.
(i). The difference between $\gamma_F^*$ for the pseudoscalar
and vector mesons, which signals the $O(\alpha a)$ systematic error.
We observed that this difference decreases as decreasing lattice
spacing, and already consistent with zero at $\beta=5.95$.
(ii). Comparison of $\gamma_F^*$ with the continuum and the Klein-Gordon
dispersion relations.
This is for a estimate of the size of $O(a^2)$ systematic uncertainty.
The results with two dispersion relations tends to
coincide with each other in decreasing the lattice spacing.
The behavior in the large quark mass region is consistent with
the expected behavior.
(iii). Response of meson masses to the change of  $\gamma_F$.
The effect of uncertainty of $\gamma_F^*$ on the meson masses is less than
1 \%, if $\gamma_F^*$ is determine at this accuracy.
This result is applicable to the relatively heavier quark mass region, 
such as $m_s < m_q$, and therefore in the region, the errors in the 
calibration is under control.

\section{Light hadron spectroscopy}
\label{sec:spectroscopy}

In this section, we apply the result of the last section
to the calculation of light hadron spectrum.
Our analysis is performed in two steps: 

(i) By taking the central value of $\gamma_F=\gamma_F^*$
we obtain the light hadron masses in 
the strange quark mass region $m_s < m_q < 2 m_s$.
By extrapolating masses in $m_{PS}^2$, 
the hadron spectrum at the physical light quark masses are
determined.
We compare our result with the result by UKQCD Collaboration 
\cite{UKQCD00}, which has been obtained on the isotropic lattice
with $O(a)$ improved quark action.

(ii) We study the response of the light hadron spectrum 
to the change of the anisotropic parameter
$\gamma_F^* \rightarrow \gamma_F^* + \delta\gamma_F$.
The extrapolation in $m_{PS}^2$ is significant to circumvent
the uncertainty in the definition of $\kappa$.

\subsection{Calculation of hadron spectrum}

The spectroscopy of light hadrons are performed on the same
lattices used in the calibration, while with smaller
numbers of configurations.
The parameters are listed in Table~\ref{tab:parameters2}.
At each $\beta$,
we use four values of $\kappa$ corresponding to the quark mass
of $m_s$--$2m_s$.
In this region, we regard that $m_q$ is sufficiently small
so that we can adopt the value of $\gamma_F^*$ in the massless limit.
Therefore the bare anisotropy is set to the central value of
$\gamma_F^*$ at $m_q=0$, which is determined in the calibration 
as the linear form in $m_q^2$.

\begin{table}[tb]
\caption{
Quark parameters used in the hadron spectroscopy.}
\begin{center}
\begin{tabular}{cccc}
\hline\hline
$\beta$ & $\gamma_F$ & values of $\kappa$ &
$N_{\mbox{conf}}$ \\
\hline
5.75& 3.909 & 0.1240, 0.1230, 0.1220, 0.1210 & 200 \\
5.95& 4.016 & 0.1245, 0.1240, 0.1235, 0.1230 & 100 \\
6.10& 4.034 & 0.1245, 0.1240, 0.1235, 0.1230 & 100 \\
\hline\hline
\end{tabular}
\end{center}
\label{tab:parameters2}
\end{table}

\begin{figure}[tb]
\centerline{\psfig{file=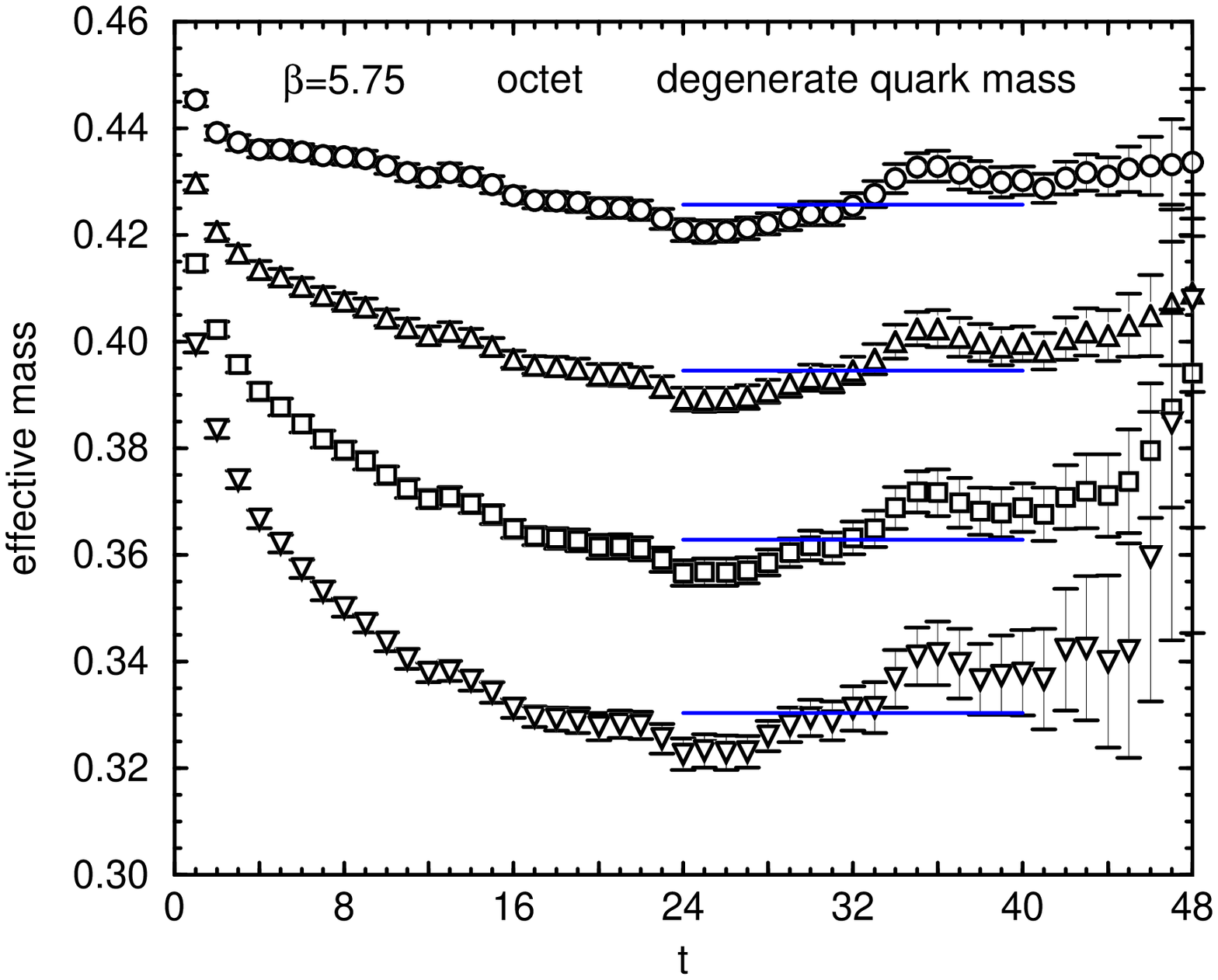,width=0.69\figwidth}
\hspace{-1.2cm}
\psfig{file=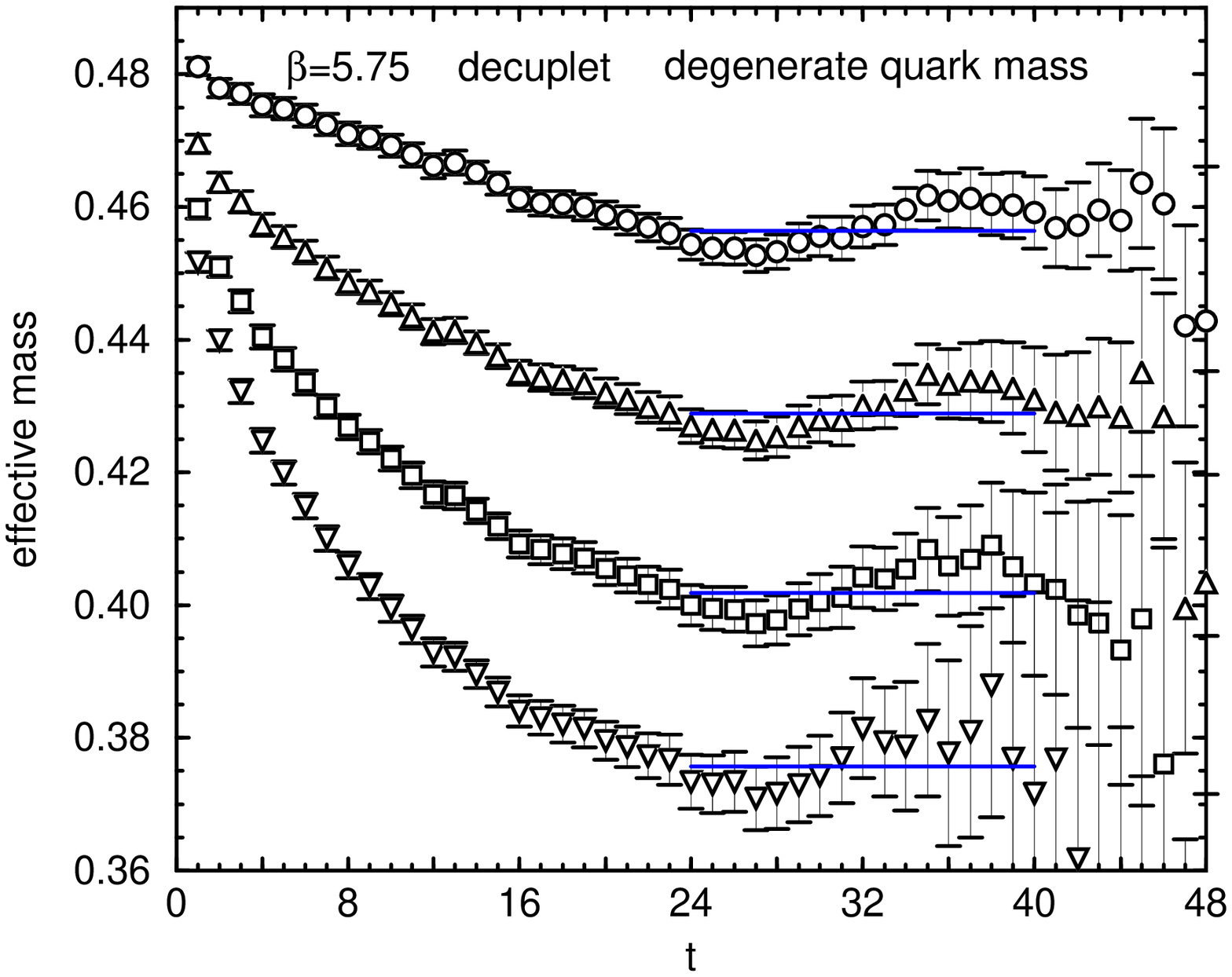,width=0.69\figwidth}}
\vspace{-0.6cm}
\caption{
Effective mass plots for octet and decuplet baryon correlators
with degenerate quark masses at $\beta=5.75$.
Horizontal solid line represent the fit range and the 
mass from the single exponential fit.}
\label{fig:ep2}
\end{figure}

We use the standard hadron operators and procedure to
extract the hadron masses.
The quark propagators are smeared at the source with Gaussian
smearing function with the deviation $\simeq 0.4$ fm,
in the Coulomb gauge.
The periodic boundary condition is adopted
in all four directions for the quark field.
For baryons, two of quarks are treated as they has degenerate masses.
Then the quark content of baryons is specified by two
$\kappa$'s, $\kappa_1$ and $\kappa_2$, for a pair of quarks
and the other quark, respectively.
Figure~\ref{fig:ep2} shows the effective mass plot for octet and
decuplet baryons with degenerate quark mass case, $\kappa_1=\kappa_2$.
The meson correlators are fitted to the single hyperbolic cosine
form.
For baryons, we apply the single exponential fit in the region
in which there is negligible contribution of the negative parity
baron from the other temporal boundary.
The result of fit is listed in Tables~\ref{tab:spectrum_E}--
\ref{tab:spectrum_G}.
For mesons, since the order of $\kappa_1$ and $\kappa_2$ is
unimportant, the masses at the exchanged set of
($\kappa_1$, $\kappa_2$) are omitted.
The masses of $\Lambda$-type octet baryon at degenerate
($\kappa_1$, $\kappa_2$) are also omitted, since they are identical
to the masses of $\Sigma$-type.

\begin{table}[tb]
\caption{
Hadron spectrum at $\beta=5.75$.
When quark masses are degenerate, i.e. $\kappa_1=\kappa_2$,
the $\Sigma$-type and the $\Lambda$-type octet baryon correlators
are identical.}
\begin{center}
\begin{tabular}{ccccccc}
\hline\hline
$\kappa_1$ & $\kappa_2$ & $m_{PS}$ & $m_{V}$ &
$m_{oct(\Sigma)}$ & $m_{oct(\Lambda)}$ & $m_{dec}$ \\
\hline
 0.1210& 0.1210& 0.22909(47)& 0.2821(10)& 0.4257(17)&     --    & 0.4564(24)\\
 0.1210& 0.1220& 0.21716(48)& 0.2730(11)& 0.4146(18)& 0.4161(18)& 0.4472(25)\\
 0.1210& 0.1230& 0.20501(51)& 0.2640(12)& 0.4034(19)& 0.4068(19)& 0.4383(27)\\
 0.1210& 0.1240& 0.19260(54)& 0.2552(13)& 0.3920(20)& 0.3977(21)& 0.4299(29)\\
\hline
 0.1220& 0.1210&    --      &     --    & 0.4059(19)& 0.4042(19)& 0.4381(27)\\
 0.1220& 0.1220& 0.20480(50)& 0.2637(12)& 0.3946(19)&    --     & 0.4289(28)\\
 0.1220& 0.1230& 0.19214(52)& 0.2546(13)& 0.3831(20)& 0.3852(21)& 0.4199(30)\\
 0.1220& 0.1240& 0.17911(55)& 0.2458(15)& 0.3713(21)& 0.3760(22)& 0.4114(33)\\
\hline
 0.1230& 0.1210&    --      &     --    & 0.3863(21)& 0.3820(20)& 0.4202(30)\\
 0.1230& 0.1220&    --      &     --    & 0.3747(22)& 0.3724(21)& 0.4109(32)\\
 0.1230& 0.1230& 0.17886(54)& 0.2454(15)& 0.3629(23)&    --     & 0.4018(35)\\
 0.1230& 0.1240& 0.16503(57)& 0.2364(17)& 0.3506(24)& 0.3538(25)& 0.3932(39)\\
\hline
 0.1240& 0.1210&    --      &     --    & 0.3674(25)& 0.3587(22)& 0.4032(37)\\
 0.1240& 0.1220&    --      &     --    & 0.3555(26)& 0.3490(24)& 0.3938(40)\\
 0.1240& 0.1230&    --      &     --    & 0.3432(27)& 0.3394(25)& 0.3845(44)\\
 0.1240& 0.1240& 0.15015(60)& 0.2272(21)& 0.3303(28)&    --     & 0.3757(51)\\
\hline\hline
\end{tabular}
\end{center}
\label{tab:spectrum_E}
\end{table}

\begin{table}[tb]
\caption{
Hadron spectrum at $\beta=5.85$.}
\begin{center}
\begin{tabular}{ccccccc}
\hline\hline
$\kappa_1$ & $\kappa_2$ & $m_{PS}$ & $m_{V}$ &
$m_{oct(\Sigma)}$ & $m_{oct(\Lambda)}$ & $m_{dec}$ \\
\hline
0.1230& 0.1230& 0.14580(40)& 0.1853(10)& 0.2788(17)&    --     & 0.3036(25)\\
0.1230& 0.1235& 0.13896(41)& 0.1804(11)& 0.2726(17)& 0.2734(18)& 0.2987(26)\\
0.1230& 0.1240& 0.13196(43)& 0.1755(11)& 0.2662(18)& 0.2680(19)& 0.2938(27)\\
0.1230& 0.1245& 0.12477(46)& 0.1707(13)& 0.2596(19)& 0.2625(20)& 0.2890(30)\\
\hline
0.1235& 0.1230&   --      &   --       & 0.2676(18)& 0.2666(18)& 0.2937(27)\\
0.1235& 0.1235& 0.13190(43)& 0.1754(11)& 0.2612(19)&    --     & 0.2888(29)\\
0.1235& 0.1240& 0.12462(45)& 0.1705(12)& 0.2546(19)& 0.2557(20)& 0.2839(31)\\
0.1235& 0.1245& 0.11709(47)& 0.1656(14)& 0.2478(20)& 0.2501(21)& 0.2790(33)\\
\hline
0.1240& 0.1230&  --       &    --      & 0.2562(20)& 0.2539(19)& 0.2839(31)\\
0.1240& 0.1235&  --       &    --      & 0.2496(21)& 0.2484(20)& 0.2789(33)\\
0.1240& 0.1240& 0.11702(47)& 0.1654(14)& 0.2427(21)&    --     & 0.2740(35)\\
0.1240& 0.1245& 0.10908(50)& 0.1605(16)& 0.2356(22)& 0.2370(23)& 0.2692(38)\\
\hline
0.1245& 0.1230&  --       &    --      & 0.2445(23)& 0.2405(21)& 0.2743(37)\\
0.1245& 0.1235&  --       &    --      & 0.2376(24)& 0.2348(22)& 0.2693(39)\\
0.1245& 0.1240&  --       &    --      & 0.2305(25)& 0.2289(24)& 0.2644(42)\\
0.1245& 0.1245& 0.10063(53)& 0.1555(19)& 0.2229(26)&   --      & 0.2597(47)\\
\hline\hline
\end{tabular}
\end{center}
\label{tab:spectrum_F}
\end{table}

\begin{table}[tb]
\caption{
Hadron spectrum at $\beta=6.10$.}
\begin{center}
\begin{tabular}{ccccccc}
\hline\hline
$\kappa_1$ & $\kappa_2$ & $m_{PS}$ & $m_{V}$ &
$m_{oct(\Sigma)}$ & $m_{oct(\Lambda)}$ & $m_{dec}$ \\
\hline
0.1230&0.1230&0.12950(29)&0.1587( 6)&0.2394(11)&    --    &0.2590(17)\\
0.1230&0.1235&0.12284(30)&0.1538( 6)&0.2332(12)&0.2340(12)&0.2542(18)\\
0.1230&0.1240&0.11603(31)&0.1491( 7)&0.2269(12)&0.2288(12)&0.2495(19)\\
0.1230&0.1245&0.10904(33)&0.1446( 8)&0.2204(13)&0.2236(13)&0.2451(21)\\
\hline
0.1235&0.1230&    --    &    --    &0.2283(12)&0.2273(12)&0.2493(19)\\
0.1235&0.1235&0.11595(31)&0.1489( 7)&0.2219(13)&    --    &0.2445(20)\\
0.1235&0.1240&0.10886(32)&0.1442( 8)&0.2154(13)&0.2166(13)&0.2399(22)\\
0.1235&0.1245&0.10153(33)&0.1398( 9)&0.2087(14)&0.2112(14)&0.2355(24)\\
\hline
0.1240&0.1230&     --    &    --    &0.2171(14)&0.2148(13)&0.2400(22)\\
0.1240&0.1235&     --    &    --    &0.2106(14)&0.2093(14)&0.2352(24)\\
0.1240&0.1240&0.10142(33)&0.1395( 8)&0.2038(14)&    --    &0.2306(26)\\
0.1240&0.1245&0.09366(35)&0.1352(10)&0.1968(15)&0.1983(16)&0.2263(30)\\
\hline
0.1245&0.1230&     --    &    --    &0.2058(16)&0.2017(15)&0.2314(28)\\
0.1245&0.1235&     --    &    --    &0.1990(16)&0.1960(15)&0.2267(30)\\
0.1245&0.1240&     --    &    --    &0.1919(17)&0.1903(16)&0.2221(34)\\
0.1245&0.1245&0.08535(36)&0.1310(12)&0.1845(18)&    --    &0.2180(41)\\
\hline\hline
\end{tabular}
\end{center}
\label{tab:spectrum_G}
\end{table}

\subsection{Extrapolation to the chiral limit}

In order to avoid the ambiguities in the definition 
of the quark mass, we extrapolate the hadron masses
to the chiral limit in terms of the pseudoscalar
meson mass squared, instead of $1/\kappa$.
We assume the relation
\begin{equation}
 m_{PS}^2(m_1, m_2) = B \cdot (m_1 + m_2),
\end{equation}
then for the degenerate quark masses, $m_1=m_2$,
$m_{PS}^2 = 2 B m_1$ holds.
Then instead of $m_i$ (i=1,2), one can use $m_{PS}(m_i,m_i)^2$
as the variable in the chiral extrapolation.
For other hadrons, vector meson and octet and decuplet baryons,
we also use the linear relations:
\begin{eqnarray}
 m_{V} (m_1, m_2) &=& m_V(0,0) + B_V \cdot (m_1 + m_2), \\
 m_{oct} (m_1, m_2, m_3) &=& m_{oct}(0,0,0)
                            + B_{oct} \cdot (m_1 + m_2 + m_3),\\
 m_{dec} (m_1, m_2, m_3) &=& m_{dec}(0,0,0)
                            + B_{dec} \cdot (m_1 + m_2 + m_3).
\end{eqnarray}
The hadron spectrum and the result of fit are shown in
Figure~\ref{fig:spectrum}.
The vertical axis is the averaged pseudoscalar meson mass squared,
\begin{equation}
 \langle m_{PS}^2(m_i) \rangle
 = \frac{1}{N_q} \sum_{i=1}^{N_q} m_{PS}^2(m_i,m_i)
 = \frac{1}{N_q} \sum_{i=1}^{N_q} 2B m_i
\end{equation}
with $N_q=2$ for mesons and $N_q=3$ for baryons.
The $\Lambda$-type baryon is not shown in the figure to avoid
that the figure become too messy.
The linear fit seems to be successful.

\subsection{Spectrum at physical quark masses}

To determine the hadron masses at the physical $u$, $d$ and $s$
quark mass, one need to set the scale of lattice.
We do not distinguish the $u$ and $d$ quark masses, and express
their mass as $m_n$.
We adopt two definitions, through the hadronic radius $r_0$,
and through the $K^*$ meson mass.
These two methods are also adopted by UKQCD Collaboration in
\cite{UKQCD00}, and convenient for comparison of our data with theirs.
In \cite{UKQCD00}, they employed the values of clover coefficient
determined in two ways: with nonperturbative renormalization
technique (NP) \cite{ALPHA97}, and tadpole-improvement (TAD)
\cite{LM93}.
Then they extrapolate the masses to the continuum limit by 
simultaneous fit of these two types of data to the linear form in
$a^2$ for NP and quadratic form in $a$ for TAD.
We compare our hadron spectrum at the physical quark masses
with the result in the continuum limit of \cite{UKQCD00},
although we ourselves do not perform the continuum extrapolation
because of lacks of sufficient number of $\beta$ as well as the
statistical accuracy.

\paragraph{Scale set by $r_0$.}

\begin{figure}[tb]
\vspace*{-0.6cm}
\centerline{\psfig{file=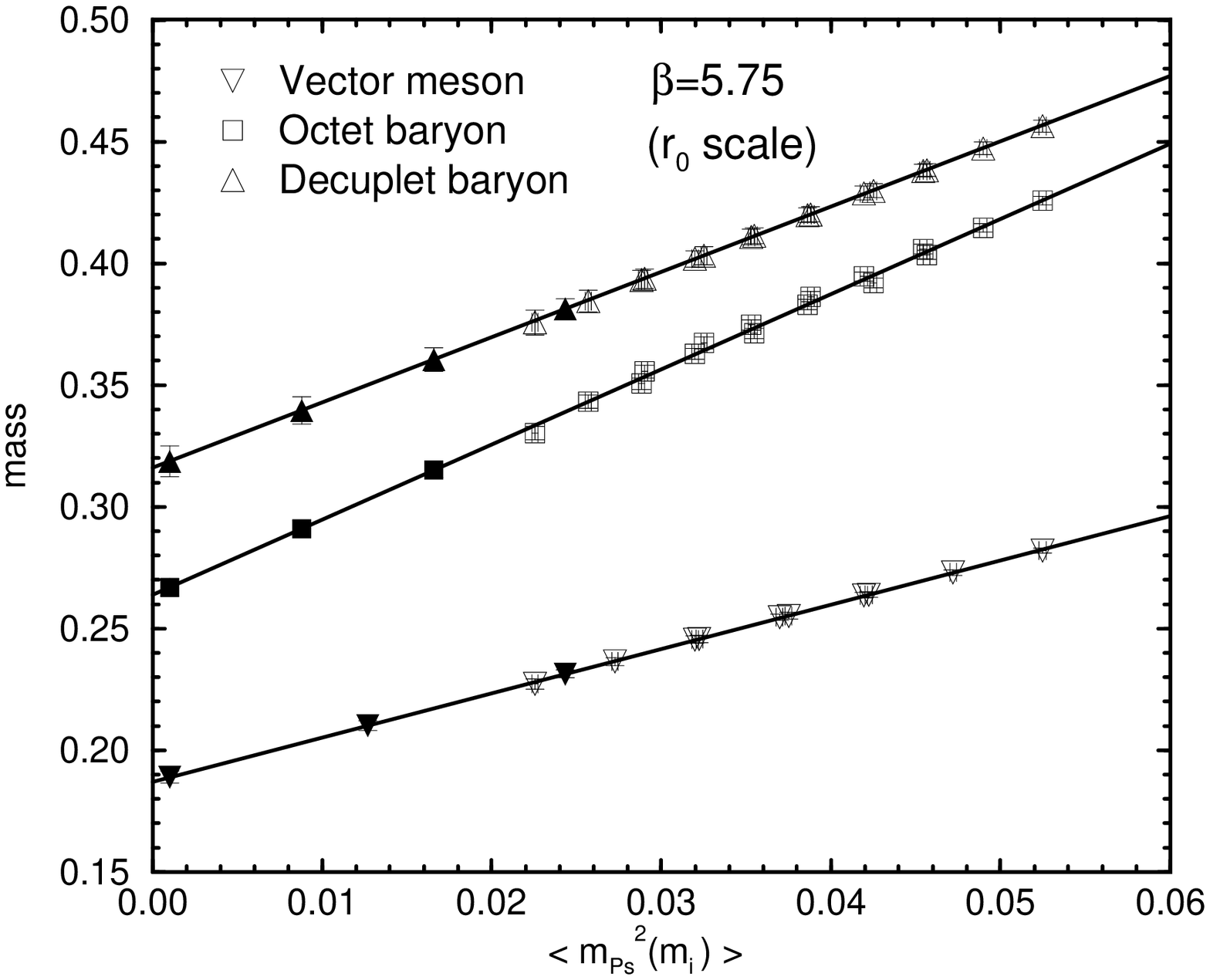,width=0.75\figwidth}}
\vspace{-1cm}
\centerline{\psfig{file=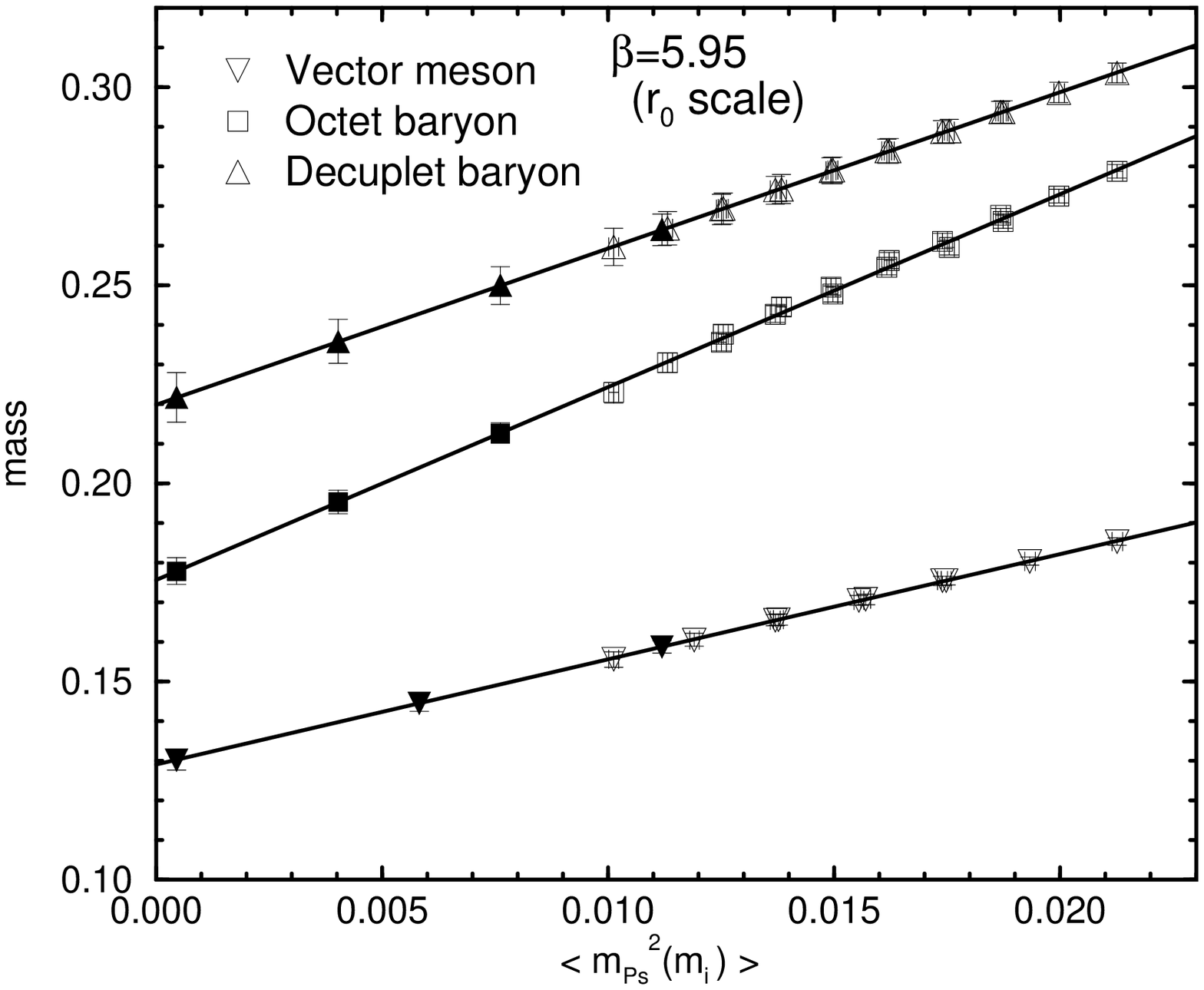,width=0.75\figwidth}}
\vspace{-1cm}
\centerline{\psfig{file=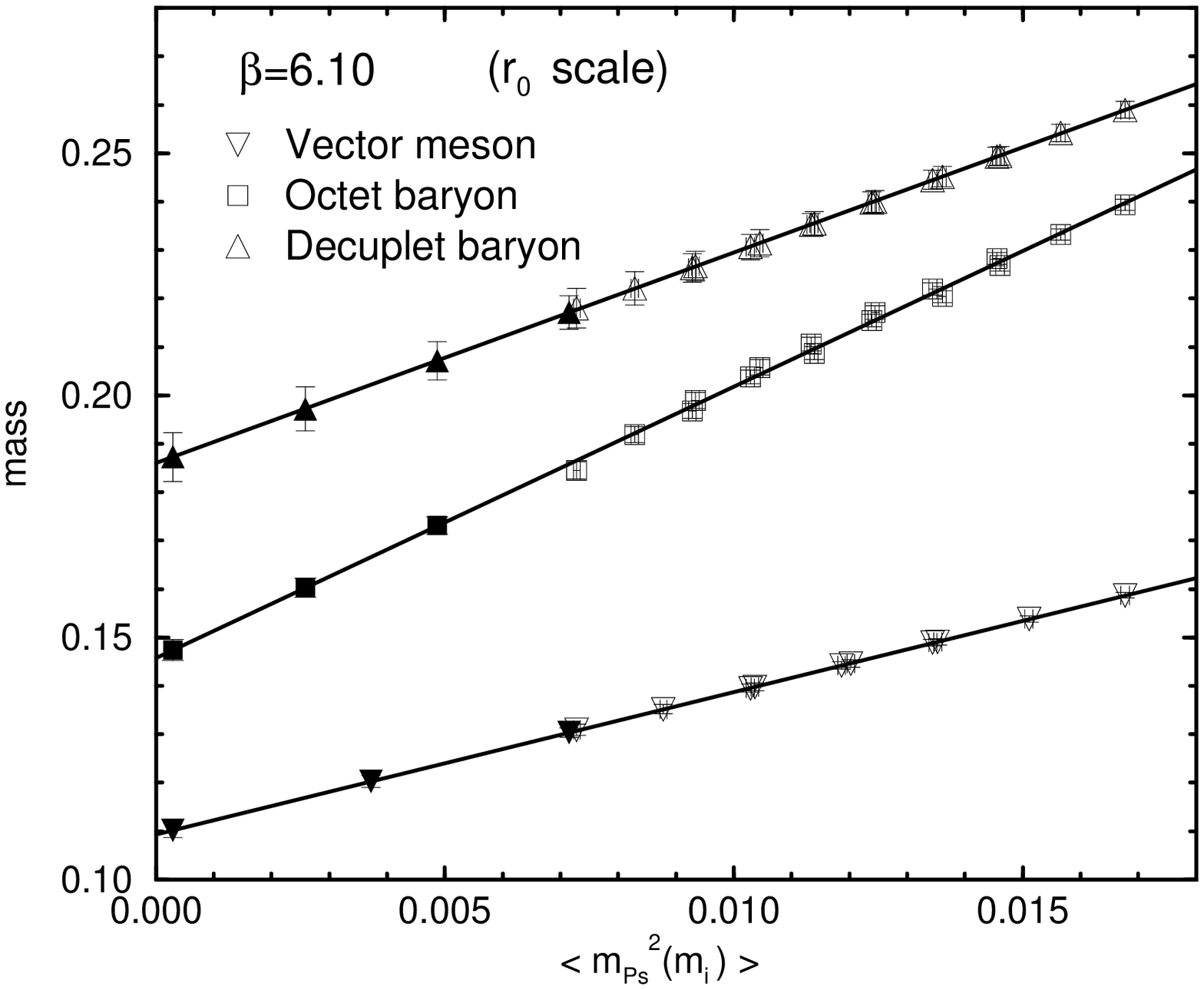,width=0.75\figwidth}}
\vspace{-0.6cm}
\caption{The masses of vector meson and the octet and decuplet
baryons together with the result of linear fits.
Only the $\Sigma$-type octet baryon is shown.
The filled symbols corresponding to the masses at the physical
$u$, $d$  and $s$ quark masses with the scale set by $r_0$.}
\label{fig:spectrum}
\end{figure}

The hadronic radius $r_0$ has been already obtained in
Section~\ref{sec:numerical}.
The corresponding values of spatial lattice cutoff are found in
Table~\ref{tab:parameters}. 
The PS meson masses squared correspond to $m_n$ and $m_s$ are then
defined with $m_{\pi}^{\pm}=139.6$ MeV and $m_K=495.7$ MeV
(isospin averaged), respectively.
These definitions are in accord with \cite{UKQCD00}.
The hadron masses extrapolated or interpolated to the physical
points are shown in Figure~\ref{fig:spectrum} and listed in
Table~\ref{tab:spectrum_phys1}.
For comparison with the result in \cite{UKQCD00},
we also list the hadron masses
multiplied by $r_0\xi$ in Table~\ref{tab:spectrum_phys1}.
In the latter case, $\xi$ appears to multiply the quantity
in the spatial lattice unit ($r_0$) to ones in the temporal lattice
unit (masses).
In our data, differences between the results at $\beta=6.10$ and
$5.95$ rather large compared with the difference between $\beta=5.95$
and $5.75$.
This would be partially due to the different $a$ dependence of
$O(\alpha a)$ and $O(a^2)$ lattice artifact, and also due to the
statistical fluctuation.
Our results of the meson masses seem to approach 
towards the continuum results by UKQCD Collaboration
on the isotropic lattices \cite{UKQCD00}.

\paragraph{Scale set by $m_{K^*}$.}

In the second case, $m_{K^*}=893.9$ MeV (isospin averaged)
is used to set the lattice scale.
First we interpolate the vector meson mass to the point 
that the ratio of PS and V meson masses is equal to the physical
value of $K^*$ and $K$ mesons.
Then this vector meson mass defines the lattice scale.
This results in the spatial lattice cutoffs  1.053(13),
1.525(27) and 1.817(22) GeV at $\beta=5.75$, $5.95$ and $6.10$,
respectively.
Then the values of $m_{PS}^2$'s correspond to the ($u$,$d$)
and $s$ quark masses are determined with experimental $K$ and $\pi$
meson masses.
The hadron masses at physical quark masses are listed in
Table~\ref{tab:spectrum_phys2}.
We observe similar tendency to the case of the scale set by $r_0$.
No signal of inconsistency with the result on isotropic lattice 
is found.

\begin{table}[tb]
\caption{
Hadron spectrum for physical quark masses with the scale set by
$r_0$.}
\begin{center}
\begin{tabular}{cccccccc}
\hline\hline
  &  & mass [GeV] &  &  &  $m \xi r_0$ & & Ref.~\cite{UKQCD00} \\
 & $\beta=5.75$ & $\beta=5.95$ & $\beta=6.10$ &
   $\beta=5.75$ & $\beta=5.95$ & $\beta=6.10$ & cont. \\
\hline
$\rho$    & 0.832(11)& 0.846(17) & 0.895(12)&
              2.105(28) & 2.141(42)& 2.265(30) & 2.35(16)\\
$K^*$     & 0.9251(90)& 0.938(13)& 0.977(10)&
              2.342(23) & 2.375(34)& 2.473(25) & 2.54(12)\\
$\phi$    & 1.0185(71)& 1.031(10)& 1.059( 8) &
              2.578(18) & 2.610(26)& 2.680(21) & 2.729(77)\\
\hline
$N$       & 1.175(16)& 1.155(22)& 1.197(18) &
              2.974(40) & 2.924(55)& 3.032(45) & 2.92(24)\\
$\Lambda$ & 1.260(13)& 1.253(19)& 1.283(16) &
              3.190(34) & 3.172(47)& 3.248(40) & 3.22(20)\\
$\Sigma$  & 1.281(14)& 1.268(19)& 1.302(16) &
              3.242(35) & 3.211(49)& 3.295(41) & 3.23(19)\\
$\Xi$     & 1.387(12)& 1.381(17)& 1.406(15) &
              3.510(31) & 3.497(43)& 3.558(37) & 3.54(15) \\
\hline
$\Delta$  & 1.403(28)& 1.440(41) & 1.521(41) &
              3.552(72) & 3.65(10) & 3.85(10)  & 3.86(37)\\
$\Sigma^*$& 1.495(25)& 1.532(36) & 1.602(37) &
              3.784(63) & 3.877(90)& 4.055(93) & 4.15(29)\\
$\Xi^*$   & 1.587(22)& 1.623(31)& 1.683(32) &
              4.017(54) & 4.109(78)& 4.260(81) & 4.44(22)\\
$\Omega$  & 1.679(18)& 1.715(26)& 1.763(28) &
              4.249(46) & 4.341(66)& 4.464(70) & 4.72(17)\\
\hline\hline
\end{tabular}
\end{center}
\label{tab:spectrum_phys1}
\end{table}

\begin{table}[tb]
\caption{
Hadron spectrum for physical quark masses with the scale set by
$m_{K^*}$.
The parameter $J$ is also quoted, while it is dimensionless quantity.}
\begin{center}
\begin{tabular}{cccccccc}
\hline\hline
  &  & mass [GeV] &  &  &  $m / m_{K^*}$ & & Ref.~\cite{UKQCD00} \\
 & $\beta=5.75$ & $\beta=5.95$ & $\beta=6.10$ &
   $\beta=5.75$ & $\beta=5.95$ & $\beta=6.10$ & cont. \\
\hline
$\rho$     & 0.796(11)& 0.795(16)&0.802(11)&
      0.891(12)& 0.890(18)& 0.898(12)& 0.921($^{+32}_{-56}$) \\
$K^*$      & 0.894(11)& 0.894(16)&0.894(11)&
                                        -- &  -- &  -- &  -- \\
$\phi$     & 0.992(12)& 0.993(16)&0.986(11)&
      1.109(13)& 1.110(18)& 1.103(12)& 1.110($^{+~8}_{-21}$)\\
\hline
$N$        & 1.125(15)&1.087(21)& 1.075(16)&
      1.259(17)& 1.216(23)& 1.202(18)& 1.14($^{+~6}_{-18}$)\\
$\Lambda$  & 1.217(14)&1.194(20)& 1.175(15)&
      1.361(15)& 1.335(22)& 1.315(17)& 1.29($^{+~5}_{-15}$)\\
$\Sigma$   & 1.236(14)&1.207(20)& 1.191(16)&
      1.382(16)& 1.351(23)& 1.332(17)& 1.29($^{+~5}_{-14}$)\\
$\Xi$      & 1.346(14)&1.328(21)& 1.307(15)&
      1.506(16)& 1.485(23)& 1.462(17)& 1.45($^{+~4}_{-10}$)\\
\hline
$\Delta$   & 1.343(27)& 1.354(39)& 1.363(37)&
      1.502(30)& 1.515(43)& 1.525(41)& 1.50($^{+17}_{-17}$)\\
$\Sigma^*$ & 1.439(25) & 1.452(35)& 1.454(33)&
      1.610(28)& 1.624(39)& 1.626(37)& 1.64($^{+13}_{-13}$)\\
$\Xi^*$    & 1.535(23)& 1.549(32)& 1.544(29)&
      1.717(25)& 1.733(36)& 1.727(32)& 1.79($^{+~9}_{-10}$)\\
$\Omega$   & 1.631(21) & 1.647(30)& 1.634(25)&
      1.825(24)& 1.842(33)& 1.828(28)& 1.93($^{+~7}_{-~8}$)\\
\hline
 $J$  & 0.3859(47) & 0.3896(95) & 0.3621(47) \\
\hline\hline
\end{tabular}
\end{center}
\label{tab:spectrum_phys2}
\end{table}

\subsection{$J$-parameter}

\begin{figure}[tb]
\centerline{\psfig{file=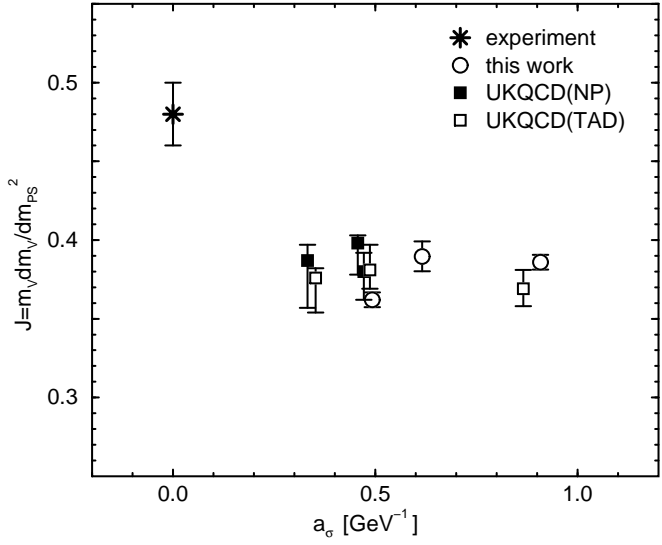,width=0.8\figwidth}}
\vspace{-0.6cm}
\caption{
The parameter $J$.
The $a_{\sigma}$ is set by using $r_0$.
Data of UKQCD Collaboration (square symbols) are taken from
Ref.~\cite{UKQCD00} on the isotropic lattices, 
and slightly horizontally shifted.}
\label{fig:J_param}
\end{figure}

The parameter $J$ was introduce to probe the quenching effect
in \cite{LM95}, and defined as
\begin{equation}
 J = m_V \left. \frac{d m_V}{dm_{PS}^2}\right|_{m_V/m_{PS}=m_{K^*}/m_K}.
\end{equation}
It is known that the quenched lattice simulation does not reproduce
the experimental value, $J=0.48(2)$, and gives about 20 \% smaller
value.
We show our result of $J$ in Figure~\ref{fig:J_param}, as the function
of lattice spacing determined by $r_0$.
We find that our results are consistent to those by UKQCD 
on the isotropic lattices in the quenched
approximation.

\subsection{Covariance of correlators}

\begin{figure}[tb]
\centerline{\psfig{file=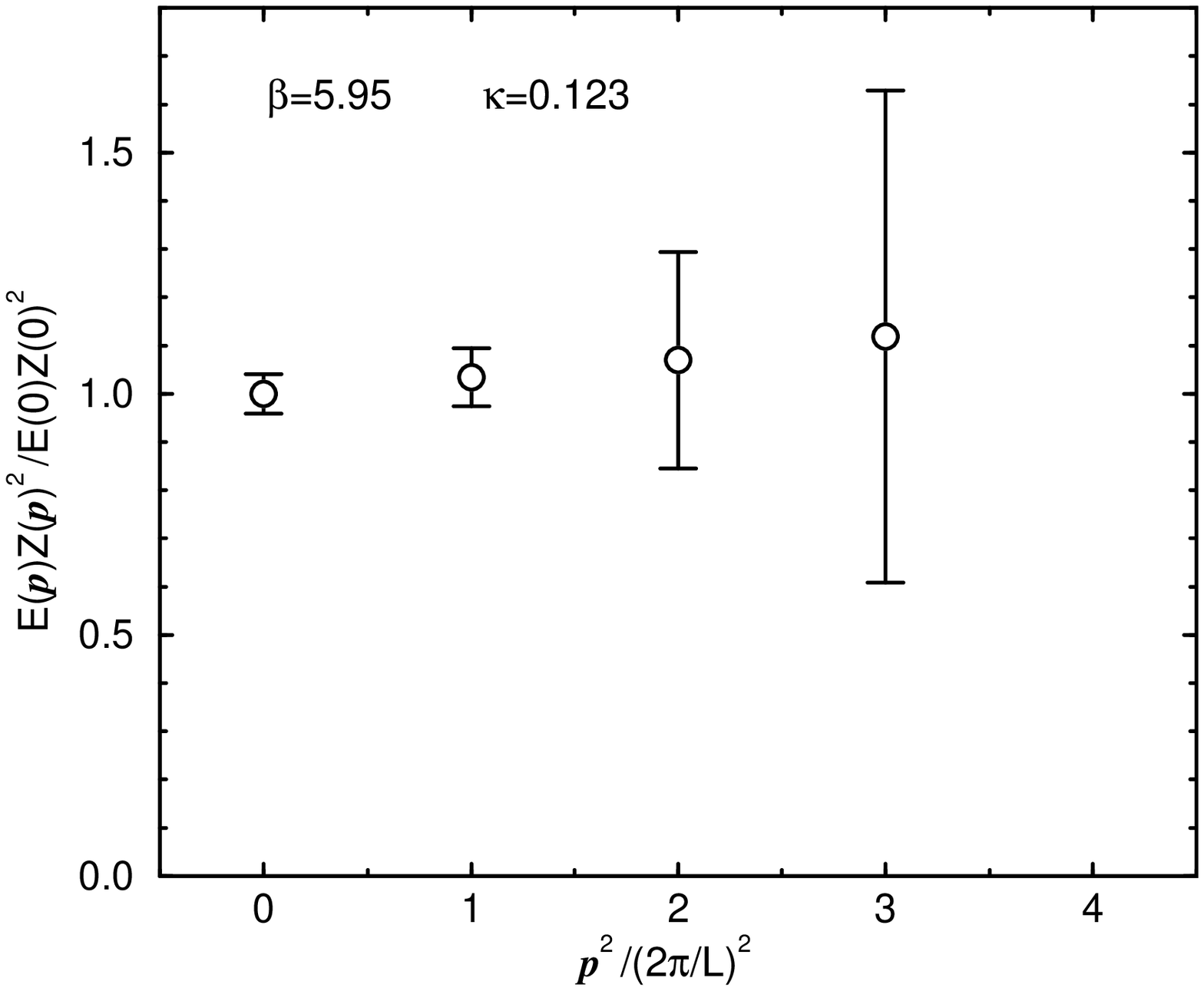,width=0.69\figwidth}
\hspace{-1.2cm}
\psfig{file=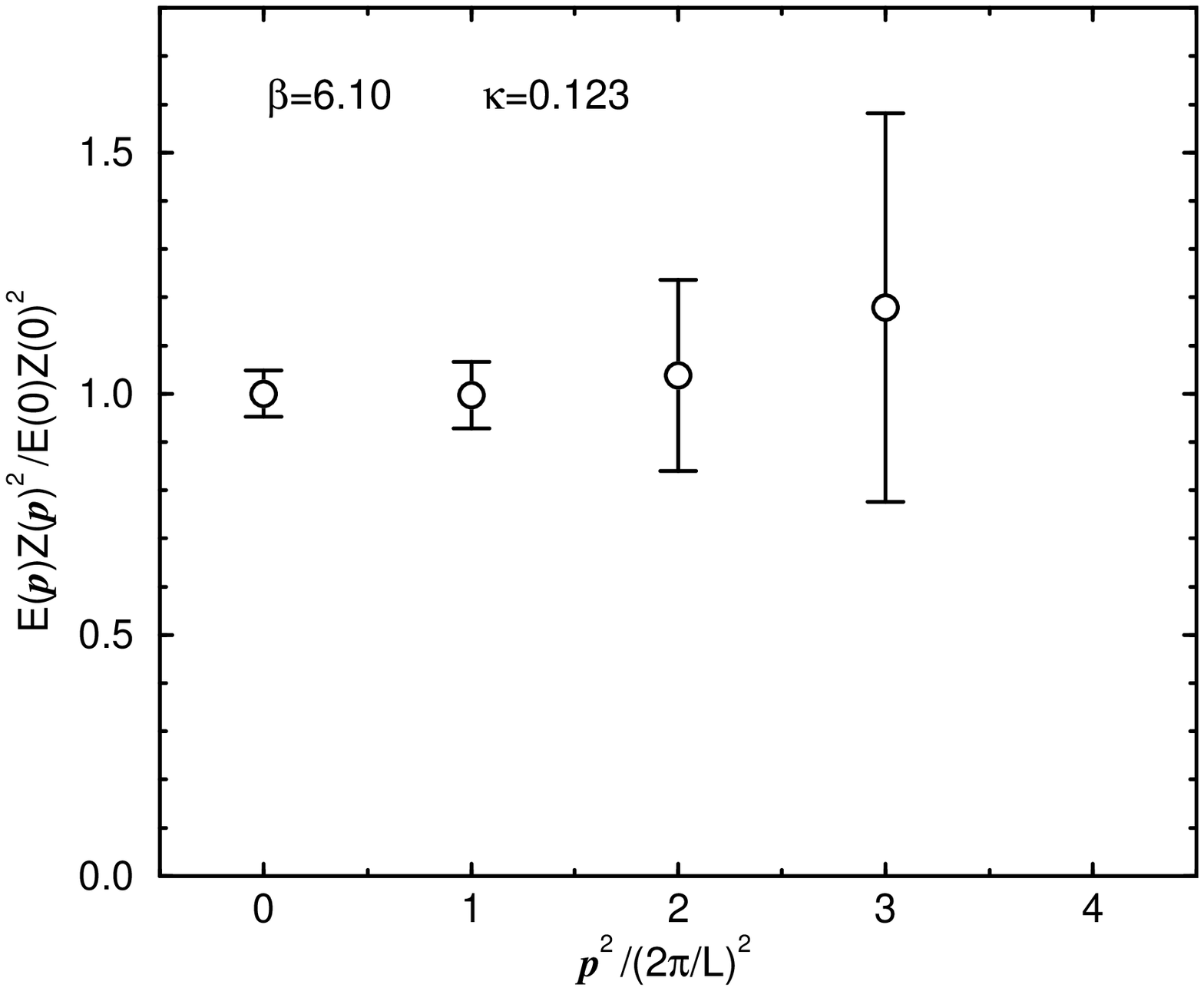,width=0.69\figwidth}}
\vspace{-0.6cm}
\caption{
The covariance of the pseudoscalar correlators.
The left figure is at $\beta=5.95$ and $\kappa=0.1230$,
and the right one is at $\beta=6.10$ and $\kappa=0.1230$.}
\label{fig:covariance}
\end{figure}

Let us consider the pseudoscalar correlator:
\begin{eqnarray}
 C_{PS}(\vec{p},t) &=& \langle O_{PS}(x) O_{PS}^{\dag}(0) \rangle 
   \nonumber \\
  &\rightarrow& \frac{Z^2(\vec{p})}{2E(\vec{p})} \exp (-E(\vec{p})t)
    \hspace{1cm} (\mbox{large } t)
\end{eqnarray}
with $Z(\vec{p})= \langle 0 | O(x) | PS(\vec{p})\rangle $.
Here we employ the covariant normalization.
For the local pseudoscalar density operator
$O(x)=\bar{q}(x)\gamma_5 q(x)$, if the
Lorentz covariance is sufficiently restored,  $Z(\vec{p})$
does not depend on the momentum $\vec{p}$.
Then
\begin{equation}
 R(\vec{p}) = \frac{E(\vec{p}) Z(\vec{p})^2 }{m_{PS} Z(0)^2}
\end{equation}
probes the restoration of covariance as the deviation from unity.
In figure~\ref{fig:covariance}, we show the momentum dependence of
$R(\vec{p})$ measured on $\beta=5.95$ and $6.10$.
At each $\beta$, the quark mass is the largest one used in the
light hadron spectroscopy.
We find that $R(\vec{p})$ at finite momentum is consistent
with $R(\vec{p}=0)$, while higher momentum states suffer from
large statistical fluctuation.
This feature is particularly important in the calculation of form
factors, in which the finite momentum states play an essential
role.

\subsection{Systematic errors of the spectrum from calibration}

\begin{figure}[tb]
\centerline{\psfig{file=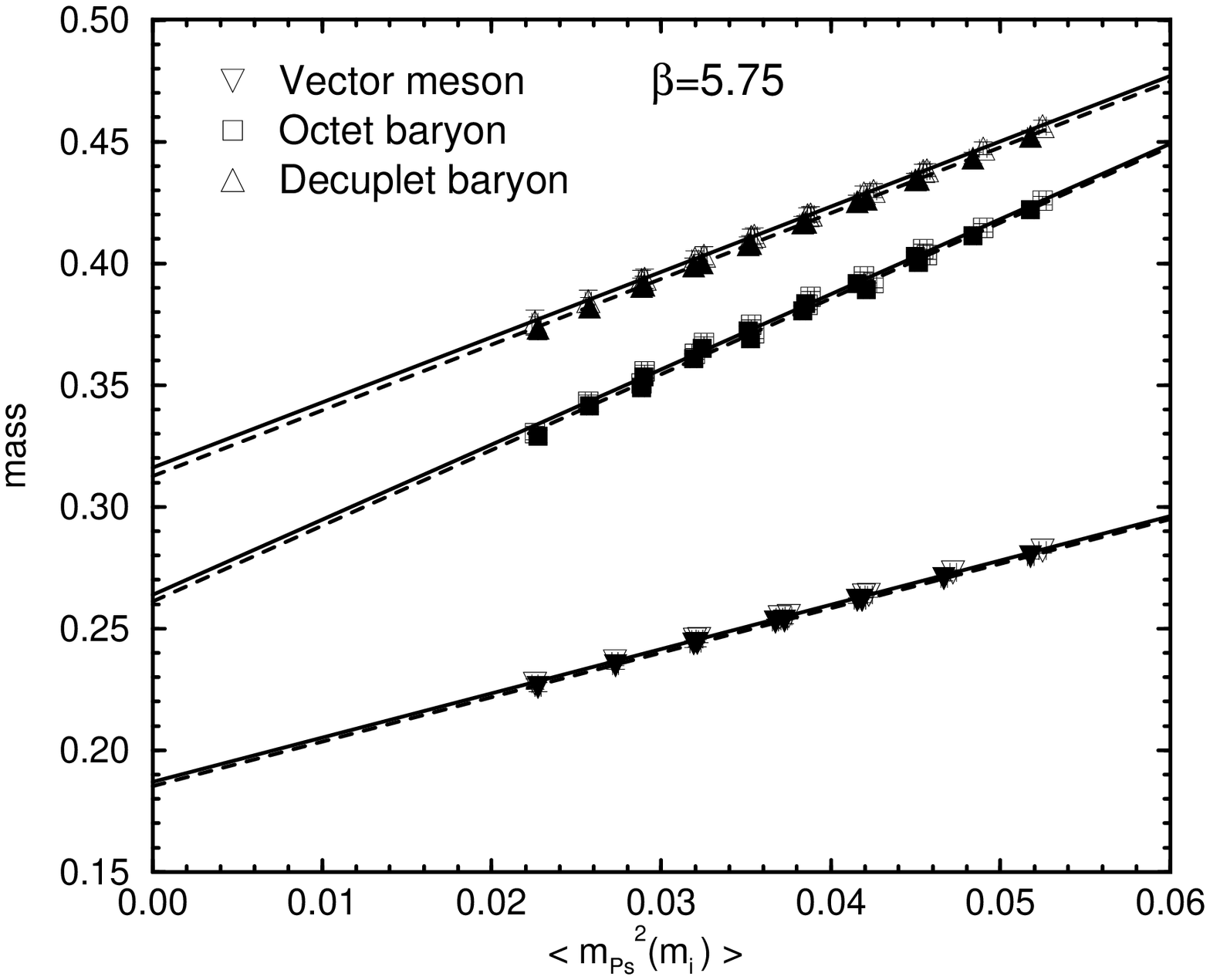,width=0.69\figwidth}
\hspace{-0.8cm}
\psfig{file=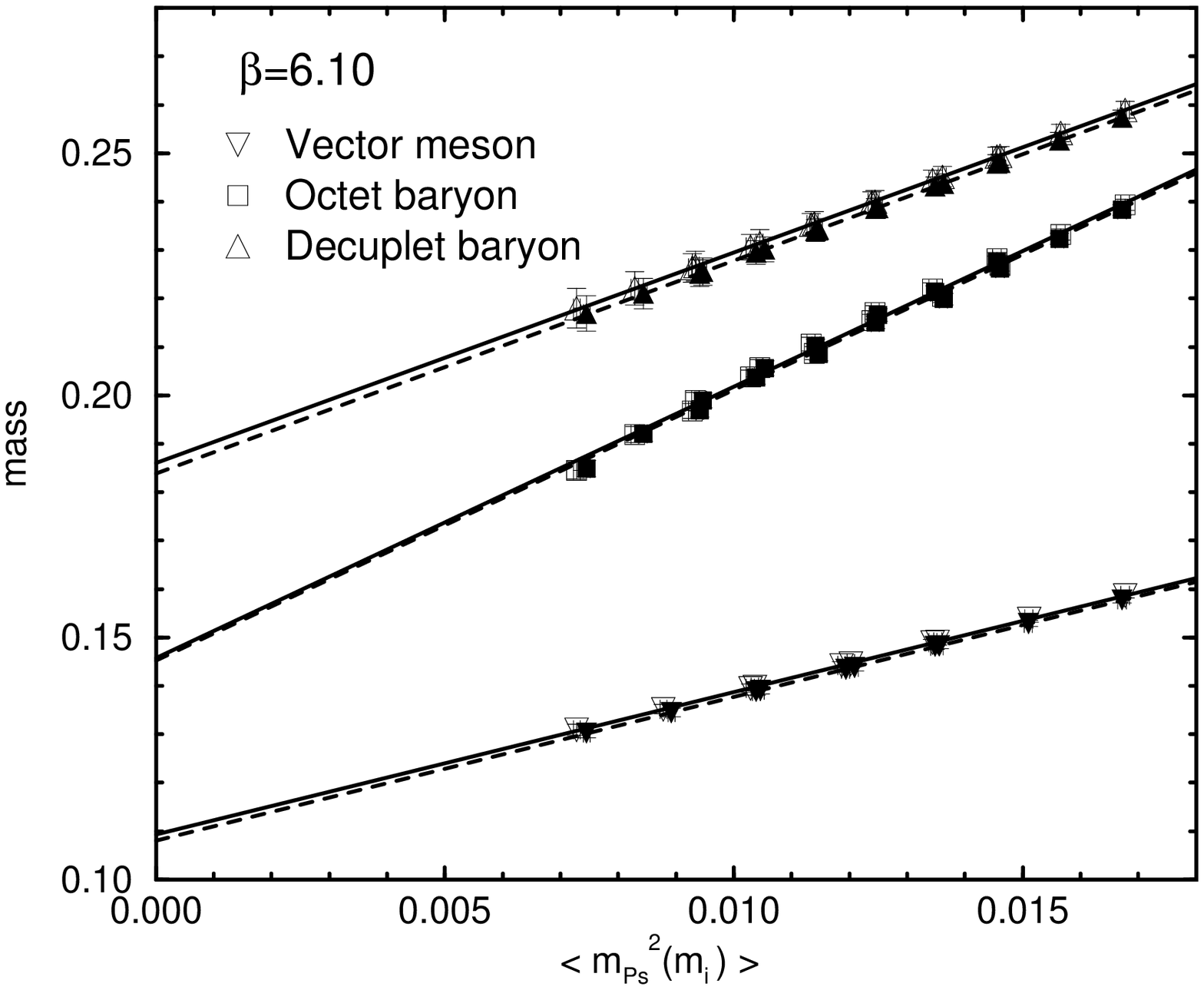,width=0.69\figwidth}}
\vspace{-0.6cm}
\caption{
The spectra with shifted $\gamma_F$ (filled symbols) together with
the results at $\gamma_F^*$ (open symbols).
The solid lines and dashed lines represent the fit results at
$\gamma_F=\gamma_F^*$ and $\gamma_F^* +\delta\gamma_F$,
respectively.}
\label{fig:spectrum2}
\end{figure}

To estimate the systematic effect due to the uncertainty of
calibration, we obtain the spectrum at the same $\kappa$'s with
slightly shifted bare anisotropy,
$\gamma_{F}' = \gamma_F^* + \delta\gamma_F$.
We set $\delta\gamma_F=0.1$, which implies about 2.5 \%
shift of bare anisotropy.
Figure~\ref{fig:spectrum2} shows the result for shifted
$\gamma_F$ together with the result for $\gamma_F^*$,
for $\beta=5.75$ and $6.10$.
There are small systematic downward shifts in the fitted lines.
The spectrum at the physical quark masses are listed in
Table~\ref{tab:spectrum_difference} as the dimensionless
combination, $m\xi r_0$ and $m/m_{K^*}$.
The difference between the masses with $\gamma_{F}'$
and $\gamma_F^*$ is slightly amplified toward the chiral limit.
Even for the lightest mass in each species,  the difference is
at most around 1 \%.
This implies that the uncertainty of hadron masses at the physical
($u$,$d$) and $s$ quark mass are about half of
uncertainty in $\gamma_F$.
With the relativistic dispersion relation, $\gamma_F^*$ at $m_q=0$
has been determined at each $\beta$ within about 2 \% ambiguity:
the statistical error of 1 \% and 
the systematic error of 1 \% in the form of fit.
Therefore there is 1 \% level uncertainty in the hadron spectrum
due to the uncertainty in calibration.
This feature would make the anisotropic lattice promising 
for future physical applications.

\begin{table}[tb]
\caption{
Spectra with shifted $\gamma_F$ in the dimensionless
combinations.
The dimensionless parameter $J$ is also listed.}
\begin{center}
\begin{tabular}{ccccccc}
\hline\hline
 & \multicolumn{3}{c}{$m\xi r_0$} &
   \multicolumn{3}{c}{$m/m_K^*$} \\
$\beta$ & 5.75 & 5.95 & 6.10 &  5.75 & 5.95 & 6.10 \\
\hline
$\rho$     &2.084(26)&2.121(40)&2.239(28)&0.891(11)&0.890(17)&0.898(11)\\
$K^*$      &2.322(21)&2.357(32)&2.448(24)&   --    &    --   &    --   \\
$\phi$     &2.560(17)&2.592(24)&2.658(19)&1.109(12)&1.110(17)&1.102(12)\\
\hline
$N$        &2.946(37)&2.913(52)&3.019(43)&1.260(16)&1.223(22)&1.212(17)\\
$\Lambda$  &3.166(32)&3.161(45)&3.236(38)&1.363(15)&1.341(21)&1.323(16)\\
$\Sigma$   &3.216(33)&3.199(46)&3.282(39)&1.383(15)&1.356(22)&1.340(17)\\
$\Xi$      &3.486(29)&3.485(41)&3.545(35)&1.507(15)&1.490(22)&1.468(17)\\
\hline
$\Delta$   &3.514(66)&3.614(96)&3.808(96)&1.503(28)&1.517(41)&1.527(38)\\
$\Sigma^*$ &3.748(58)&3.847(85)&4.014(85)&1.610(26)&1.625(37)&1.627(34)\\
$\Xi^*$    &3.983(50)&4.080(73)&4.221(75)&1.717(24)&1.734(34)&1.728(30)\\
$\Omega$   &4.217(43)&4.313(62)&4.427(65)&1.824(23)&1.842(32)&1.829(27)\\
 \hline
 $J$       &         &         &         &
             0.3838(43) & 0.3871(86) & 0.3614(42) \\
\hline\hline
\end{tabular}
\end{center}
\label{tab:spectrum_difference}
\end{table}

\section{Conclusion}
  \label{sec:conclusion}

In this paper, we studied the $O(a)$ improved quark action
on the anisotropic lattice with anisotropy $\xi=a_{\sigma}/a_{\tau}=4$.
The bare anisotropy $\gamma_F^*$, with which $\xi_F=\xi$ holds,
is determined for the whole quark mass region below the charm quark mass,
including the chiral limit in 1 \% statistical accuracy.
In the massless limit, there is also about 1 \% systematic uncertainty
in extrapolating $\gamma_F^*$ to $m_q=0$.

The uncertainties in the calibration due to the discretization
errors are studied by changing the physical inputs or conditions. 
(i) We have shown that 
the dispersion relations for the pseudoscalar and the vector mesons
give values of $\gamma_F^*$ which differ by 1\% at $\beta=5.75$,
while they show no difference at $\beta=5.95$ and $6.10$.
(ii) Two different choices of the lattice dispersion relations,
namely the naive continuum form and the Klein-Gordon form, 
also lead to the results which differ by 3\% for $\beta=5.75$,
but we found no difference at $\beta=6.1$ with $m_q < 0.2 a^{-1}_{\tau}$.
These systematic uncertainties tend to vanish toward the continuum limit.

The light hadron spectrum was studied using the central value
of tuned bare anisotropy, $\gamma_F^*(m_q=0)$.
We found that it is consistent with the result on the isotropic lattice
by UKQCD Collaboration.
It was found that a change of $\gamma_F^*$ by 2 \%
would lead to a change of the spectrum by 1 \%
for the physical quark masses.
We also investigated the Lorentz invariance of the matrix element
of the pseudoscalar operator as a consistency check.

The main disadvantage in using the anisotropic lattice
would lie in the additional systematic uncertainty caused by
the calibration.
There are two types of errors in $\gamma_F^*$.
The first type consists of the statistical error and the error in the chiral
extrapolation, which was estimated to be 2 \% level.
The second type consists of $O(\alpha a)$ and $O(a^2)$ 
systematic uncertainties,
which was estimated to be 4\% at $\beta=5.75$ and smaller for larger
$\beta$. In total, there is 6 \% ambiguity at $\beta=5.75$, which 
corresponds to our coarsest lattice, and 2--3 \% ambiguities at $\beta=6.10$.
The relative errors in hadron spectrum are half of those in
$\gamma_F^*$.  Since the contribution from the second types
of error vanishes in the continuum limit, we expect to obtain the 
hadron spectrum to 1 \% accuracy in the continuum limit.
This result is encouraging for further applications.
The anisotropic lattice would already be applicable 
to quantitative studies which requires a few percent accuracy.
To achieve higher accuracy, nonperturbative tuning of 
the clover coefficients are required.

Since the range of quark masses where the systematic errors 
are under control covers the charm
quark region, it is also important to apply the present anisotropic
lattice simulation to the charmonium and D meson systems.

\section*{Acknowledgments}

We thank J.~Harada, A.S.~Kronfeld, O.~Miyamura, N.~Nakajima,
Y.~Nemoto, H.~Suganuma and T.T.~Takahashi
for useful discussions.
The simulation has been done on
NEC SX-5 at Research Center for Nuclear Physics, Osaka University and
Hitachi SR8000 at KEK (High Energy Accelerator Research Organization).
H.~M. is supported by Japan Society for the Promotion of Science
for Young Scientists.
H.~M. was also supported in the early stage of this work
by the center-of-excellence (COE) program at RCNP, Osaka University.
T.~O. is supported by the Grant-in-Aid of the Ministry
of Education No. 12640279.
T.~U. is supported by the center-of-excellence (COE) program
at CCP, Tsukuba University.

\end{document}